\documentclass[%galley,%
11pt,tightenlines,%
nofootinbib,
preprintnumbers,%
showpacs,%
amsfonts,%
prd]{revtex4}

\usepackage{graphicx}
\newcommand{\be}{\begin{equation}}
\newcommand{\ee}{\end{equation}}
\newcommand{\bea}{\begin{eqnarray}}
\newcommand{\eea}{\end{eqnarray}}

\newcommand{\0}{\over }

\newcommand{\6}{\partial }
\newcommand{\g}{g_{\rm eff}}
\newcommand{\vac}{{\rm vac}}
\newcommand{\muMS}{\bar\mu_{\rm MS}}

\newcommand{\im}{{\rm Im}\,}

\newcommand{\real}{\mathrm{Re}\,}
\newcommand{\imag}{\mathrm{Im}\,}

\newcommand{\cS}{\mathcal{S}}
\newcommand{\cN}{\mathcal{N}}

\newcommand{\Tr}{\mathrm{Tr}\,}

\begin{document}

\preprint{TUW-04-12}

\title{Non-Fermi-Liquid Specific Heat of Normal Degenerate Quark Matter}
\date{\today}
\author{A. Gerhold}
\affiliation{Institut f\"ur Theoretische Physik, Technische
Universit\"at Wien, \\Wiedner Haupstr.~8-10, 
A-1040 Vienna, Austria }
\author{A. Ipp}
\affiliation{Institut f\"ur Theoretische Physik, Technische
Universit\"at Wien, \\Wiedner Haupstr.~8-10, 
A-1040 Vienna, Austria }
\author{A. Rebhan}
\affiliation{Institut f\"ur Theoretische Physik, Technische
Universit\"at Wien, \\Wiedner Haupstr.~8-10, 
A-1040 Vienna, Austria }
\begin{abstract}
We compute the low-temperature behavior of the specific heat of
normal (non-color-superconducting)
degenerate quark matter as well as that of an ultradegenerate electron gas.
Long-range magnetic interactions lead to non-Fermi-liquid behavior
with an anomalous leading $T\ln T^{-1}$ term.
Depending on the thermodynamic potential used as starting point, this effect
appears as a consequence of the logarithmic singularity in the
fermion self-energy at the Fermi surface or directly as a
contribution from the only weakly screened quasistatic magnetic
gauge bosons. We show that a calculation of Boyanovsky and
de Vega claiming the absence of a leading $T\ln T^{-1}$ term
missed it by omitting vector boson contributions to the internal energy.
Using a formulation which collects all nonanalytic contributions
in bosonic ring diagrams, we systematically calculate
corrections beyond the well-known leading-log approximation.
The higher-order terms of the low-temperature
expansion turn out to also involve fractional powers 
$T^{(3+2n)/3}$ and we explicitly
determine their coefficients up to and including 
order $T^{7/3}$ as well as the subsequent logarithmically
enhanced term $T^3 \ln (c/T)$. 
We derive also a hard-dense-loop resummed expression which contains
the infinite series of anomalous terms to leading order in the coupling and
which we evaluate numerically.
At low temperatures,
the resulting deviation of the specific heat from
its value in naive perturbation theory is significant in the
case of strongly coupled normal quark matter and thus of potential relevance
for the cooling rates of (proto-)neutron stars with a quark matter component.
\end{abstract}
\maketitle

\section{Introduction}

It is well-known that long-range magnetic interactions in
a degenerate electron gas lead to non-Fermi-liquid behavior
which manifests itself in 
the appearance of an anomalous contribution to the 
low-temperature limit of entropy and specific
heat proportional to $\alpha T\ln T^{-1}$ as discovered by
Holstein, Norton, and Pincus
\cite{Holstein:1973} over thirty years ago. 
While this effect is perhaps too small for experimental
detection in nonrelativistic situations, it drew renewed theoretical attention more recently
\cite{Reizer:1989,Gan:1993,Chakravarty:1995}
after the detection of non-Fermi-liquid behavior in
%resistivity and specific heat of
the normal state of high-temperature superconductors \cite{Varma:1989}
and in other systems of strongly correlated electrons,
which may be due to effective gauge field dynamics
(see also \cite{Polchinski:1992ed,Polchinski:1994ii,Nayak:1994ng}).

In deconfined degenerate quark matter, the analogous effect
can more easily be important because the larger
coupling constant $\alpha_s$ together with the relatively 
large number of gauge bosons
increases the numerical value of the effect by orders of magnitude.
In contrast to the case of a high-temperature quark-gluon plasma,
chromomagnetostatic fields are expected to remain unscreened in the
low-temperature limit \cite{Son:1998uk} and thus lead to the same singularities
in the fermion self-energy that are responsible for the
breakdown of the Fermi-liquid description in the nonrelativistic
electron gas considered in \cite{Holstein:1973}.

An important consequence of such non-Fermi-liquid behavior in
quantum chromodynamics (QCD) is a reduction of the magnitude of
the gap in color superconductors \cite{Son:1998uk,Brown:1999aq,%Brown:2000eh,%
Wang:2001aq} which on the basis of weak-coupling calculations
are estimated to have a critical temperature 
in the range between 6 and 60 MeV \cite{Rischke:2003mt}.
Quark matter above this temperature, and unpaired quark matter component
also below it, has long-range chromomagnetic interactions that
should lead to an anomalous specific heat with possible
relevance for the cooling of young neutron stars 
as pointed out by Boyanovsky and de Vega 
\cite{Boyanovsky:2000bc,Boyanovsky:2000zj}.
However, in Ref.~\cite{Boyanovsky:2000zj} these authors
claimed that the $\alpha T \ln T^{-1}$ term in the specific heat
as reported in \cite{Holstein:1973,Gan:1993,Chakravarty:1995}
would not exist, neither in QCD nor in QED.
Instead they obtained a $\alpha T^3 \ln T$ correction
to the leading ideal-gas behavior, which by renormalization-group
arguments was resummed into a $T^{3+O(\alpha)}$ correction
as the leading non-Fermi-liquid effect on the specific heat.\footnote{%
Resummation of the $\alpha T \ln T^{-1}$ term along the
lines of Ref.~\cite{Boyanovsky:2000zj} would have
led to a $T^{1+O(\alpha)}$ term instead.}
At low temperatures, such a contribution would be rather
negligible compared to standard perturbative corrections to
the ideal-gas result $\propto T$.

In a numerical study of the exactly solvable large-flavor-number limit
of QCD and QED \cite{Moore:2002md%,Ipp:2003zr
} at nonzero chemical
potential \cite{Ipp:2003jy}, two of us however found
that the entropy at low temperature has a behavior suggestive of a
$\alpha T \ln T^{-1}$ term. In Ref.~\cite{Ipp:2003cj}, the three of us
have recently reproduced the known $\alpha T \ln T^{-1}$ term in
entropy and specific heat, together
with further anomalous higher-order corrections,  
in an analytical calculation that should apply equally to the
case of finite flavor number.
This calculation is however organized in a form which does not
allow one to compare directly with the calculation
of Ref.~\cite{Boyanovsky:2000zj} where all
$\alpha T \ln T^{-1}$ terms appeared to cancel.

In this paper we shall therefore investigate the approach
of Ref.~\cite{Boyanovsky:2000zj}, which derived the specific
heat from a formula for the internal energy, and compare
with two somewhat more direct calculations, one using a self-consistent
formula for the entropy and another using an expression
for the thermodynamic potential that becomes exact in the limit
of large flavor number.

As we shall demonstrate, all these approaches agree eventually
and do give a leading $\alpha T \ln T^{-1}$ term for the specific heat.
In the calculation using a self-consistent
formula for the entropy (Sect.~\ref{sectII}) 
the $\alpha T \ln T^{-1}$ term arises
as a contribution from the spectral density of the fermions
with their logarithmic singularity in the self-energy.
There are also $\alpha T \ln T^{-1}$ 
contributions from the gauge boson sector, but
these cancel in the end, which thus validates the
(in our opinion not unquestionable)
starting point of the original calculation by Holstein et al.
\cite{Holstein:1973}.
On the other hand, in the calculation of the specific heat from the internal
energy (Sect.~\ref{sectIII}) 
we find that keeping only the fermionic contributions
leads to a cancellation of the leading $\alpha T \ln T^{-1}$ term,
just as observed in Ref.~\cite{Boyanovsky:2000zj}. However, it turns
out that in this approach
the contribution of the gauge bosons to the specific heat cannot be
neglected, but now contains the complete leading logarithm.

In Sect.~\ref{sectIV} we describe the details of a calculation
which %in contrast to the former approaches 
allows us to systematically
go beyond the leading-log approximation. Besides completing the
argument of the leading logarithm, we find fractional powers 
$T^{(3+2n)/3}$ and we determine their coefficients up to and including 
order $T^{7/3}$ as well as the subsequent logarithmically
enhanced term $T^3 \ln (c/T)$.
This low-temperature expansion requires that the temperature
is much smaller than the scale set by the Debye mass.
At temperatures of the order of the Debye mass or larger, but
still much smaller than the quark chemical potential, a complete
leading-order result 
which contains the infinite series of anomalous terms 
is obtained in Sect.~\ref{sect4plus}. It
involves a hard-dense-loop resummed one-loop expression, which
we evaluate numerically in Sect.~\ref{sectV}.
This allows us to study the quality of the low-temperature
expansion, and to compare with
the exact results for the large-flavor-number limit.
For the sake of this comparison we shall throughout use the notation
\begin{equation}\label{geffdef}
\g^2 = \left\{
        \begin{array}{cc} 
        \displaystyle \frac{g^2 N_f}{2} \, , & {\rm QCD} \, , \\ & \\
        g^2 N_f \, , & {\rm QED} \, . \\ \end{array} \right.
\end{equation}
with $g$ the coupling constant and $N_f$ the number of
quark (or electron) flavors. 
At finite $N_f$ we finally evaluate our results numerically for
a range of coupling which may be relevant for (normal) quark matter
in (proto-) neutron stars, with the finding that there is an
interesting range of temperature where the anomalous specific
heat exceeds significantly the ideal-gas value.

\section{Anomalous specific heat from the entropy}
\label{sectII}

The specific heat $\mathcal C_v$ per unit volume is
defined as the logarithmic derivative of the entropy density
with respect
to temperature at constant volume and number density
\be
\mathcal C_v=T\left({\6{\cal S}\0\6T}\right)_{\cal N}.
\ee

This is related to derivatives of the thermodynamic potential
with respect to $T$ and $\mu$ by \cite{LL:V-Cv}
\be
C_{v}=T\left\{ \left( \frac{\partial {\cal S}}{\partial T}\right) _{\mu }-{\left( \frac{\partial \mathcal N}{\partial T}\right) ^{2}_{\mu }}{\left( \frac{\partial \mathcal N}{\partial \mu }\right)^{-1} _{T}}\right\},
\ee
but at low temperatures one has
\begin{equation}
  \mathcal C_v=T\left(\partial \cS\over\partial T\right)_\mu+\mathcal{O}(T^3), \label{x1}
\end{equation}
so that both $\mathcal C_v$ and $\mathcal S$ contain the same $T\ln T$ term, if any.

The entropy as first derivative of the thermodynamic potential $\Omega$
with respect to $T$ is in some important respects a simpler quantity
than $\Omega$.
In gauge theories with fermions the latter
%The thermodynamic potential for QCD 
is given by the following functional of the full 
propagators $D$ (for gauge bosons), $S$ (for fermions)
\cite{Luttinger:1960ua},
\begin{equation}
  \beta\Omega[D,S]={1\over2}\Tr\ln D^{-1}-{1\over2}\Tr\Pi D-\Tr\ln S^{-1}
  +\Tr\Sigma S+\Phi[D,S],
\end{equation}
%we shall take into account only the 2-loop contribution $\Phi^{(2)}$. 
where $\Phi$ is a series of 2-particle-irreducible (skeleton) diagrams
and where for simplicity we assumed a ghost-free gauge.

Using the fact that $\Omega[D,S]$ is stationary with respect to
variations of $D$ and $S$, one can derive an expression
for the entropy which to two-loop order in the skeleton expansion
is entirely given by propagators and self-energies 
\cite{Vanderheyden:1998ph,Blaizot:1999ip%,Blaizot:2000fc
}.
Neglecting the longitudinal gluon mode, and the antiparticle 
contributions in the fermionic sector, this reads
\begin{eqnarray} 
  &&\!\!\!\!\!\!\!\!\!\cS=\left(\partial P\over\partial T\right)_\mu\simeq-\int {d^4K\over(2\pi)^4}
  \Bigg[2N_g{\partial n_b(\omega)\over\partial T}
  \left(\imag\ln D_T^{-1}-\imag\Pi_T\,\real D_T\right)\nonumber\\
  &&\qquad\qquad\qquad+4NN_f{\partial n_f(\omega)\over\partial T}
  \left(\imag\ln S_+^{-1}+\imag\Sigma_+\,\real S_+\right)\Bigg]+\cS^\prime, \label{x2}
\end{eqnarray}
where $D_T^{-1}=-\omega^2+k^2+\Pi_T$, $S_+^{-1}=-\omega+k-\Sigma_+$,
$n_b(\omega)=(e^{\omega/T}-1)^{-1}$, and $n_f(\omega)=(e^{(\omega-\mu)/T}+1)^{-1}$.

In the original derivation of the anomalous specific heat in QED
by Holstein et al. \cite{Holstein:1973}, only the term
involving $\imag\ln S_+^{-1}$ in the quark part had been taken into account,
by way of reference to a formula by Luttinger \cite{Luttinger:1960}
(Eq.~(46) therein), which
is fully justified actually only for standard Fermi-liquid systems.

Since the calculation of Ref.~\cite{Holstein:1973} has been
questioned by Boyanovsky and de Vega 
\cite{Boyanovsky:2000zj}, we consider the more general expression
above. For the moment
we are interested only in
terms of order $g^2T\ln T$, so we can
neglect $\cS^\prime$, which vanishes 
at two-loop order in the skeleton expansion \cite{Vanderheyden:1998ph}
and should therefore only give contributions which are suppressed by an 
additional factor of $g^2$.

\subsection{Quark part}
In Eq. (\ref{x2}) we have the following contribution from the quarks,
\begin{eqnarray}
  &&\cS_{(\mathrm{q})}=-4NN_f\int{d^4K\over(2\pi)^4}
  {\partial n_f(\omega)\over\partial T}\nonumber\\
  &&\qquad\qquad\times\bigg[\imag\ln\left(-\omega+k-\Sigma_+\right)
  +\imag\Sigma_+\,\real{1\over-\omega+k-\Sigma_+}\bigg]\nonumber\\
  &&\,\qquad\simeq-{1\over\pi^3}NN_f\int_0^{\infty}dk\,k^2\int_{-\infty}^\infty d\omega
  {\partial n_f(\omega)\over\partial T}\nonumber\\
  &&\qquad\qquad\times\bigg[\imag\ln\left(-\omega+k\right)
  -\real\Sigma_+\,\imag{1\over-\omega+k}\bigg],\quad \label{x9}
\end{eqnarray}
where we have performed an expansion with respect to $\Sigma_+$, keeping only the free term
and the term corresponding to a single quark self energy insertion\footnote{%Higher self energy insertions
%would produce terms which are suppressed at least with an additional factor of $g^2\ln(M/T)$.}. 
Note that, diagrammatically, the part with a single self energy insertion corresponds to the 
gluon ring diagram of section \ref{sect22}.}.
The free term gives the particle 
contribution to the free fermionic entropy density,
\begin{equation}
  \cS_{(\mathrm{q})}^{\mathrm{free}}\simeq NN_f{\mu^2 T\over3}.
\end{equation}
In the last term in Eq. (\ref{x9})
the factor $\imag\,1/(-\omega+k)$ forces the self energy to be on the mass shell.
Using the expression for $\Sigma_+$ given in 
\cite{Brown:1999aq,Manuel:2000mk}, 
\begin{equation}
  \Sigma_+\simeq {g^2C_f\over24\pi^2}(\omega-\mu)\ln\left({M^2\over(\omega-\mu)^2}\right)
  +i{g^2C_f\over12\pi}|\omega-\mu|,\label{x21}
\end{equation}
which is nonanalytic in $\omega$ 
(but not with respect to $k$ \cite{Brown:1999aq}),
we obtain
\begin{equation}
  \cS_{(\mathrm{q})}^{NLO}={N_g\over\pi^2}\int_0^{\infty}dk\,k^2
  {\partial n_f(k)\over\partial T}
  {\g^2\over24\pi^2}(k-\mu)\ln{M^2\over(k-\mu)^2}.
\end{equation} 
With the substitution $k=Tz+\mu$ we find that 
the integral is dominated by small values of $z$, 
and therefore we may send the lower integration limit
to $-\infty$. Then we obtain at order $T\ln T$
\begin{equation}
  \cS_{(\mathrm{q})}^{NLO}
  ={\g^2N_g\mu^2 T\over36\pi^2}\ln\left({M\over T}\right). \label{x23}
\end{equation}

This result agrees with the one of Holstein et al. \cite{Holstein:1973}
after correcting a factor of $4$ therein, as done previously in
Ref.~\cite{Chakravarty:1995}.

\subsection{Gluon part}\label{sect22}

The gluon part $\cS_{({\rm g})}$ is given by the first line of Eq. (\ref{x2}). Using the
relation %\cite{Blaizot:2000fc}
\begin{equation}
  \imag\ln D^{-1}=\arctan\left({\imag\Pi\over\real D^{-1}}\right)-\pi\epsilon(\omega)
  \theta\left(-\real D^{-1}\right),
\end{equation}
we write $\cS_{({\rm g})}=\cS_{(\mathrm{cut})}+\cS_{(\Pi)}+\cS_{(\mathrm{pole})}$,
with
\begin{eqnarray}
  &&\cS_{(\mathrm{cut})}=2N_g\int {d^4K\over(2\pi)^4}{\partial n_b(\omega)\over\partial T}
  \arctan\left({\imag\Pi_T\over\omega^2-k^2-\real\Pi_T}\right),\\
  &&\cS_{(\Pi)}=-2N_g\int {d^4K\over(2\pi)^4}{\partial n_b(\omega)\over\partial T}
  \imag\Pi_T\,\real{1\over\omega^2-k^2-\Pi_T},\\
  &&\cS_{(\mathrm{pole})}=2N_g\int {d^4K\over(2\pi)^4}{\partial n_b(\omega)\over\partial T}
  \pi\epsilon(\omega)\theta\left(\omega^2-k^2-\real\Pi_T\right),
\end{eqnarray}
where we again neglect the contribution of the Debye-screened
longitudinal gluons.
For the cut term we use the approximation $\omega\ll k$, because it can be checked that
including terms of higher order in $\omega$ would only produce terms of higher order than
$T\ln T$ (see section \ref{sectIV}). In this region we have
\begin{equation}
  \Pi_T\simeq-i{\g^2\mu^2\omega\over 4\pi k}.
\end{equation}
Introducing an UV-cutoff $k_c$ for the moment, we obtain
\begin{equation}
  \cS_{(\mathrm{cut})}\simeq{N_g\over 2\pi^3}\int_0^{k_c}dk\,k^2\int_{-\infty}^\infty d\omega 
  {\partial n_b(\omega)\over\partial T}
  \arctan\left({\g^2\mu^2\omega\over4\pi k^3}\right).
\end{equation}
%with $m^2=\g^2\mu^2/(2\pi^2)$.
In order to evaluate this integral we make the substitution $y=\omega/T$, $x=4\pi k^3/(\g^2\mu^2T)$.
Keeping only the term of order $T\ln T$, we obtain the cutoff-independent result
\begin{equation}
  \cS_{(\mathrm{cut})}\simeq{\g^2N_g\mu^2T\over 36\pi^2}\ln\left(M^\prime\over T\right). \label{x7}
\end{equation}
The determination of the constant $M^\prime$ 
requires a more accurate calculation and will be carried out in Sect.~\ref{sectIV}.

Next we evaluate $\cS_{(\Pi)}$. 
Following similar steps as in the computation of 
$\cS_{(\mathrm{cut})}$, we find 
\begin{eqnarray}
  &&\cS_{(\Pi)}\simeq2N_g\int_0^{k_c}{dk\,k^2\over2\pi^2}\int_{-\infty}^\infty{d\omega\over2\pi}
  {\partial n_b(\omega)\over\partial T}
  \left(-{\g^2\mu^2\omega\over4\pi k}\right){k^2\over k^4+
  \left({\g^2\mu^2\omega\over4\pi k}\right)^2}\nonumber\\
  &&\qquad\simeq-{\g^2N_g\mu^2T\over 36\pi^2}\ln\left(M^\prime\over T\right).
\end{eqnarray}
We observe that at order $T\ln T$ this expression just cancels the contribution from Eq. (\ref{x7}).

Finally we consider the pole part. To leading order,
i.e. in the hard dense loop (HDL) approximation 
\cite{Braaten:1990mz,Altherr:1992mf,Vija:1995is,Manuel:1996td}, 
we have at low temperature
$\mu\,{\partial\Pi_T / \partial\mu}\simeq2\Pi_T$, and therefore
\begin{eqnarray}
  &&\mu{\partial\cS_{(\mathrm{pole})}\over\partial\mu}=-2N_g\int {d^4K\over(2\pi)^4}
  {\partial n_b(\omega)\over\partial T}
  \pi\epsilon(\omega)\,\delta\left(\omega^2-k^2-\real\Pi_T\right)2\,\real\Pi_T\nonumber\\
  &&\qquad\qquad=-4\pi N_g\int {d^4K\over(2\pi)^4}
  {\partial n_b(\omega)\over\partial T}
  (\omega^2-k^2)\epsilon(\omega)\,\delta\left(\omega^2-k^2-\real\Pi_T\right),\qquad
\end{eqnarray}
where we have discarded contributions $\sim T^3$
which are negligible in the low-temperature limit.
Using \cite{LeB:TFT}
\begin{equation}
  \epsilon(\omega)\,\delta\left(\real D_T^{-1}\right)
  =Z_T(k)\left[\,\delta(\omega-\omega_T(k))-\delta(\omega+\omega_T(k))\,\right]
\end{equation}
we find
\begin{equation}
  \mu{\partial\cS_{(\mathrm{pole})}\over\partial\mu}=-{2N_g\over\pi^2}\int_0^\infty dk\,k^2
  {\partial n_b(\omega_T(k))\over\partial T}(\omega_T(k)^2-k^2)Z_T(k).
\end{equation}
We can estimate this integral as follows.
Assuming $T\ll\omega_p\propto \g\mu$, we have the inequalities
\begin{eqnarray}
  &&\int_0^\infty dk\,k^4{\partial n_b(\omega_T(k))\over\partial T}Z_T(k)<
  \int_0^\infty dk\,k^4{\partial n_b(\omega_T(k))\over\partial T}{1\over2k}\nonumber\\
  &&\qquad<{1\over2}\int_0^{\omega_p} dk\,k^3{\partial n_b(\omega_p)\over\partial T}+
  {1\over2}\int_{\omega_p}^\infty dk\,k^3{\partial n_b(k)\over\partial T}
  \simeq {\omega_p^5\over8T^2}e^{-\omega_p/T}\qquad\qquad \label{i17}
\end{eqnarray}
and
\begin{eqnarray}
  &&\int_0^\infty dk\,k^2{\partial n_b(\omega_T(k))\over\partial T}\omega_T(k)^2Z_T(k)<
  \int_0^\infty dk\,k^2{\partial n_b(\omega_T(k))\over\partial T}(\omega_p+k)^2{1\over2k}\nonumber\\
  &&\qquad<{1\over2}\int_0^{\omega_p} dk\,k(\omega_p+k)^2{\partial n_b(\omega_p)\over\partial T}+
  {1\over2}\int_{\omega_p}^\infty dk\,k(\omega_p+k)^2{\partial n_b(k)\over\partial T}\nonumber\\
  &&\qquad\simeq {17\omega_p^5\over24T^2}e^{-\omega_p/T}. \label{i18}
\end{eqnarray}
Apart from
terms $\sim T^3$ which are dropped in the derivative with respect to $\mu$,
this crude estimate (which we shall refine in Sect.~\ref{sect4plus}
below) shows that the pole contribution
is exponentially suppressed, essentially because
of $\omega_T\ge\omega_p$.

\subsection{Result}

In total we find the following result for the entropy at low temperature,
\begin{equation}
  \cS=\cS_{(\mathrm{g})}+\cS_{(\mathrm{q})}\simeq NN_f{\mu^2T\over3}+{\g^2N_g\mu^2 T\over36\pi^2}
  \ln\left({M\over T}\right). \label{x13}
\end{equation}

From Eq. (\ref{x7}) we see that the $T\ln T$ term can also be obtained by starting only from the expression
\begin{equation}\label{S2gl}
  \cS\simeq NN_f{\mu^2T\over3}-2N_g\int {d^4K\over(2\pi)^4}
  {\partial n_b(\omega)\over\partial T}
  \imag\ln D_T^{-1},
\end{equation}
with $D$ the resummed gluon propagator. This formula
corresponds to integrating out the fermions,
as has indeed been done in the approach 
of Ref.~\cite{Gan:1993}. %,Ipp:2003cj}. 

On the other hand, we see from Eqs. (\ref{x9}) and (\ref{x23}) that one
equally gets the correct result by using only the purely fermionic expression
\begin{equation}
  \cS\simeq-4NN_f\int {d^4K\over(2\pi)^4}{\partial n_f(\omega)\over\partial T}
  \imag\ln (-\omega+k-\real\Sigma_+),
\end{equation}
which justifies the starting point of Refs.\
\cite{Holstein:1973,Chakravarty:1995}.

\section{Specific heat from the energy density}
\label{sectIII}

In Ref.~\cite{Boyanovsky:2000zj}, which did not find a term
$g^2 T \ln T^{-1}$ in the specific heat, the starting point was
instead the internal energy density.

The energy density can be obtained from the expectation value of the energy momentum tensor,
\begin{equation}
  \mathcal{U}={1\over V}\int d^3x\,\langle T^{00}(x)\rangle,
\end{equation}
and the specific heat is then given by
\begin{equation}
  \mathcal C_v=\left({d\mathcal{U}\over dT}\right)_\mathcal{N}. \label{cvu}
\end{equation}
Here the temperature derivative has to be taken at constant particle number density, in contrast with 
the calculation of the low temperature specific heat in the previous section, 
where all temperature derivatives were 
taken at constant chemical potential, 
see Eq. (\ref{x2}). 
In \cite{Boyanovsky:2000zj} this fact was mentioned as explanation
for the disagreement with the previous calculation, but, as 
Eq. (\ref{x1}) makes clear, this could only affect terms of
order $T^3$ in the low-temperature expansion.
Indeed, we shall
show now that 
%We shall now demonstrate that 
%both methods ultimately yield the same result.
a complete calculation based on the internal energy also
leads to a $g^2 T \ln T^{-1}$ in the specific heat.

In \cite{Boyanovsky:2000zj} the specific heat is computed using the following formula for the
total energy density,
\begin{equation}
  \mathcal{U}=2\int d\omega\int {d^3k\over(2\pi)^3}n_f(\omega)\,\omega\,\rho_+(\omega,k), 
\end{equation}
where $\rho_+$ is the spectral density of the positive energy component of the quark propagator
(see below). It should be noted that this formula is incorrect even for a theory with only
instantaneous interactions of the type
\begin{equation}
  H_{int}={1\over2}\int d^3x\,d^3x^\prime\,\psi_\alpha^\dag({\bf x} t)\psi_\beta^\dag({\bf x^\prime} t)
  V_{\alpha\alpha^\prime,\beta\beta^\prime}({\bf x}-{\bf x^\prime})\psi_{\beta^\prime}({\bf x^\prime} t)
  \psi_{\alpha^\prime}({\bf x} t),
\end{equation}
in which case the correct formula reads \cite{FetW:Q}
\begin{equation}
  \mathcal{U}=2\int d\omega\int {d^3k\over(2\pi)^3}n_f(\omega)\,{1\over2}(\omega+k)\rho_+(\omega,k).
  \label{u1}
\end{equation}
The anomalous behaviour of the specific heat comes from dynamically screened interactions, 
whose non-instantaneous character cannot be neglected.
%plays a decisive role for  physical properties,
%for instance for the color superconductivity gap \cite{Son:1998uk}. 
It might be
rather difficult to generalize Eq. (\ref{u1}) directly for non-instantaneous interactions, because
one would have to use an effective Hamiltonian which is nonlocal in time. Therefore, we
will use the full energy momentum tensor of QCD without
integrating out the gluons.

The energy momentum tensor can be written as a sum of three distinct pieces,
\begin{equation}
  T^{\mu\nu}=T^{\mu\nu}_{(\mathrm{q})}+T^{\mu\nu}_{(\rm{g})}+T^{\mu\nu}_{(\mathrm{int.})},
  %+T^{\mu\nu}_{(\mathrm{HDL})}
%  +T^{\mu\nu}_{(\mathrm{e.o.m.})},
\end{equation}
corresponding to the quark part, the gluon part, 
%the HDL part related to soft gluons,
and the interaction part.
%and a part which is proportional to the equations of motion.
The contributions of these
parts will be evaluated in the following subsections. We will neglect
gluon self interactions and ghost contributions, since they give only higher order corrections at low temperatures.

\subsection{Quark part}
The quark part is given by
\begin{equation}
  T^{00}_{(\mathrm{q})}=i\sum_{f}\bar\psi\gamma^0\partial^0\psi, \label{f1}
\end{equation}
where we have written explicitly the sum over flavor space.
This is the (only) contribution which is taken into account by Boyanovsky and
de Vega \cite{Boyanovsky:2000zj}.
We now repeat their calculation, 
but for simplicity without the renormalization group improvement
of the quark propagator proposed in \cite{Boyanovsky:2000zj}. 
Taking into account only the positive energy component of the quark propagator,
we find
\begin{equation}
  \mathcal{U}_{(\mathrm{q})}=
  2NN_f\int d\omega\int {d^3k\over(2\pi)^3}n_f(\omega)\,\omega\,\rho_+(\omega,k), 
  \label{e7}
\end{equation}
where the spectral density is defined as $\rho_+\equiv{1\over\pi}\imag S_+$.

In order to obtain the specific heat from Eq. (\ref{cvu}), we have to determine first the temperature
dependence of the chemical potential from the condition
\begin{equation}
  {d\mathcal{N}\over dT}\equiv{\partial\mathcal{N}\over\partial T}
  +{d\mu\over dT}{\partial\mathcal{N}\over\partial\mu}=0, \label{e10}
\end{equation}
where the particle number density $\mathcal N$ is given by
\begin{equation}
  \mathcal{N}=2NN_f\int d\omega\int {d^3k\over(2\pi)^3}n_f(\omega)\,\rho_+(\omega,k) \label{e11}
\end{equation}
(up to anti-particle contributions.) We expand $\cN$ with respect to $g$,
\begin{equation}
  \cN=\cN_0+\g^2\cN_2+\ldots
\end{equation} 
The free contribution $\cN_0$ is given by
\begin{equation}
  \cN_0=NN_f\left({\mu^3\over3\pi^2}+{\mu T^2\over3}\right). 
\end{equation}
In $\cN_2$ we are only interested in contributions which contain $\ln(M/T)$. 
Such terms arise from infrared singularities caused by the transverse
gluon propagator, 
which are dynamically screened. This
corresponds to scattering processes of quarks which are close to the Fermi surface. Therefore the anomalous
terms come from the region $k\sim\omega\sim\mu$, where $\Sigma_+$ is given by (\ref{x21}).
Subtracting the temperature independent part, we find then
\begin{equation}
  \g^2\cN_2=2NN_f\int d\omega\int {d^3k\over(2\pi)^3}
  \left(n_f(\omega)-\theta(\mu-\omega)\right)\delta(\omega-k+\real\Sigma_+)\bigg|_{\mathcal{O}(\g^2)},
  \label{x40}
\end{equation}
where we have approximated the spectral density by a delta function, since the imaginary part of
$\Sigma_+$ turns out to be 
negligible compared to its real part. The integration can be performed easily, with the result 
\begin{equation}
  \g^2\cN_2\simeq{\g^2N_g\mu T^2\over 36\pi^2}\ln\left(M\over T\right).
\end{equation}
We notice that this result is consistent with the result for the entropy, Eq. (\ref{x13}).
Now we can solve Eq. (\ref{e10}) at low temperature, 
\begin{equation}
  {d\mu\over dT}=-{2\pi^2T\over3\mu}-
{g^2C_fT\over18\mu}\ln\left(M\over T\right).
  \label{x33}
\end{equation}
The approximate solution to this differential equation is given by
\begin{equation}
  \mu(T)\simeq\mu(0)\left(1-{\pi^2T^2\over3\mu(0)^2}-{g^2C_fT^2\over36\mu(0)^2}
  \ln\left({M\over T}\right)\right). \label{x34}
\end{equation}
Eqs. (\ref{x33}) and (\ref{x34}) correctly reproduce the beginning 
of the perturbative expansions of the corresponding formulae in 
\cite{Boyanovsky:2000zj} (Eqs. (2.37), (2.38)).

For the specific heat we obtain from Eq. (\ref{e7}), following the same steps as in the calculation of
$d\mathcal{N}/dT$,
\begin{equation}
  C_{v(\mathrm{q})}\simeq NN_f\mu^2T+NN_f{\mu^3\over\pi^2}{d\mu\over dT}+{\g^2N_g\mu^2T\over18\pi^2}
  \ln\left({M\over T}\right). \label{x36}
\end{equation}
Using Eqs. (\ref{x33}) and (\ref{x34}) we find that the $T\ln T$-terms cancel,
\begin{equation}
   C_{v(\mathrm{q})}\simeq NN_f{\mu^2T\over3} + O(T^3),
\end{equation}
as stated in \cite{Boyanovsky:2000zj}. We should emphasize that this cancellation has nothing to do with the
nonperturbative renormalization group method which is employed in \cite{Boyanovsky:2000zj} (and which has recently been criticized in 
Ref.~\cite{Schafer:2004zf}).

The authors of Ref.~\cite{Boyanovsky:2000zj} also determined a contribution
which prior to the renormalization group improvement corresponds to
a term of order $g^2 T^3 \ln(M/T)$. This type of nonanalytic terms
however appears already in regular Fermi-liquids \cite{Carneiro:1975}
and moreover is subleading to ordinary perturbative corrections
$g^2 \mu^2 T$ at low $T$, 
which have not been evaluated in \cite{Boyanovsky:2000zj}.

\subsection{Gluon part}

We now turn to the gluon part of the energy density, which has
explicitly been neglected in Ref.~\cite{Boyanovsky:2000zj}.
This is given by
\begin{equation}
  T^{00}_{({\rm g})}={1\over2}\left({\bf E}^a\cdot{\bf E}^a+{\bf B}^a\cdot{\bf B}^a\right).
\end{equation}
Neglecting gluon self interactions, and keeping only the transverse part of the gluon propagator, we obtain
\begin{equation}
  \mathcal{U}_{({\rm g})}\simeq2N_g\int{d^4K\over(2\pi)^4}n_b(\omega)\,\imag\left(\left(\omega^2+k^2\right)D_T\right).
  \label{e51}
\end{equation}
The pole contribution to this integral is again exponentially
suppressed, therefore we only have to consider the cut contribution. 
At low temperature the temperature dependence of the
gluon self energy can be neglected and 
%is almost temperature independent.
%Subtracting the non-$n_b$-part 
%(which is less IR-singular than the $n_b$-part), 
we find
\begin{equation}
  \mathcal{U}_{({\rm g})}\simeq
  2N_g\int{d^3K\over(2\pi)^3}
\int_0^\infty {d\omega\0\pi} n_b(\omega)(\omega^2+k^2)
  {{\g^2\mu^2\omega\over 4\pi k}\over\left(\omega^2-k^2\right)^2
  +\left({\g^2\mu^2\omega\over 4\pi k}\right)^2}, \label{u52}
\end{equation}
where we have dropped less infrared-sensitive contributions
not involving the Bose distribution.
After the substitution $\omega=Ty$, $k^3=\g^2\mu^2Tx/(4\pi)$ the integral can be readily done, with the result
\begin{equation}
  \mathcal{U}_{({\rm g})}\simeq{\g^2N_g\mu^2T^2\over72\pi^2}\ln(M/T),
\end{equation}
which gives the following contribution to the specific heat at order $T\ln T$,
\begin{equation}
  C_{v({\rm g})}\simeq{\g^2N_g\mu^2T\over36\pi^2}\ln(M/T). \label{e52}
\end{equation}
Again the determination of the constant $M$ would require a more accurate calculation, similar
to the one in section \ref{sectIV}.

\subsection{Interaction part}
The interaction part is given by
\begin{equation}
  T^{00}_{(\mathrm{int})}=g\sum_f\bar\psi\gamma^0A^0_aT_a\psi. 
\end{equation}
The expectation value of this term is essentially given by the
resummed gluon ring diagram as in \cite{Moore:2002md}. However, here only the longitudinal component of 
the gluon propagator appears in the loop%\footnote{The transverse projection operator 
%$\mathcal{A}^{\mu\nu}$ is orthogonal to the other projection operators $\mathcal{B}^{\mu\nu}$,
%$\mathcal{C}^{\mu\nu}$ and $\mathcal{D}^{\mu\nu}$. Therefore a longitudinal gluon can 
%(of course) never transmute itself into a transverse one as it propagates.}
. This mode is subject to Debye screening, so 
it can contribute only to the normal Fermi liquid part of the specific heat.

%\subsection{Equations of motion}
%In deriving the expression for the energy momentum tensor,
%one sometimes adds or subtracts terms which vanish by imposing the equations of motion, for instance
%\begin{equation}
%  T^{00}_{(\mathrm{e.o.m.})}=-\sum_f\bar\psi (i\Ds+\mu\gamma_0)\psi.
%\end{equation}
%Using the Schwinger-Dyson equation one can show that the corresponding contribution to the energy
%density is given by
%\begin{equation}
%  \mathcal{U}_{(\mathrm{e.o.m.})}=
%  -{1\over V}\sum_f\int d^3x\,\langle\bar\psi (i\Ds+\mu\gamma_0) \psi\rangle=NN_f\delta^4(0). \label{e14}
%\end{equation}
%This is a formally divergent contribution to the vacuum energy density. But since the result
%(\ref{e14}) is temperature independent, it does not contribute to the specific heat.

%%In an analogous way one can show that adding a term of the form
%%$A_\nu^a\,\delta S/\delta A^{\mu a}$ to the energy momentum tensor
%%does not contribute to the specific heat. 

\subsection{Result}

We have thus found that the only contribution to the specific heat at
order $T\ln T$ when calculated along the lines of Ref.~\cite{Boyanovsky:2000zj}
comes from the gluon part, Eq. (\ref{e52}). While this
confirms the observation of Ref.~\cite{Boyanovsky:2000zj} that
the quark contribution of order $g^2 T\ln T$ cancels against
a similar term in the temperature dependence of the chemical
potential at fixed number density, it shows that the (explicit) neglect of
the gluon contribution to the internal energy in
\cite{Boyanovsky:2000zj} is not justified.

Curiously enough, the gluon contributions can be neglected when
calculating the anomalous specific heat from the entropy functional
(\ref{x2}). But the calculation of section \ref{sectII} can also
be viewed as receiving a net contribution only from
a purely gluonic term (\ref{S2gl}), since the anomalous contribution
contained in the $S_{(\Pi)}$ part is equal but opposite in sign
to the fermionic part $S_{(q)}$. In this respect the calculation
based on the internal energy is perfectly in line with the
calculation based on the entropy.
The different possibilities for organizing the calculation of
the anomalous contribution to the specific heat thus correspond
to `integrating out' first the fermionic degrees of freedom
or first the gluonic ones.

\section{Higher orders in the low-temperature specific heat}
\label{sectIV}

In this section we shall evaluate higher terms in the low-temperature 
expansion of the specific heat which go
beyond the leading log approximation. 
%In particular we shall aim at determining all contributions
%up to and including order $T^3 \ln (c/T)$.
A convenient starting point is the following expression for
the pressure, 
which can be viewed as 
the result of having integrated out first the fermionic degrees
of freedom, and which thus concentrates on the effects of the
only dynamically screened transverse gauge bosons in the
low-temperature expansion of the thermodynamic potential:
%which becomes exact in the limit of large flavor number $N_f$, and which at finite
%$N_f$ has an error of order $g^4\mu^4$ in the regime $T\ll g\mu$,
\begin{eqnarray}
  &&P=NN_f\left({\mu^4\over12\pi^2}+{\mu^2T^2\over6}+{7\pi^2T^4\over180}\right)\nonumber\\
  &&\quad-N_g\int {d^3q\over(2\pi)^3}\int_0^\infty {dq_0\over\pi}\bigg[2\left([n_b+{1\over2}]\imag\ln D_T^{-1}
  -{1\over2}\imag\ln D_{\rm vac}^{-1}\right)\nonumber\\
  &&\qquad+\left([n_b+{1\over2}]\imag\ln{D_L^{-1}\over q^2-q_0^2}
  -{1\over2}\imag\ln{D_{\rm vac}^{-1}\over q^2-q_0^2}\right)\bigg]+
+O(g^2 T^4)
+{O}(g^4\mu^4).   
%\quad (T\ll g\mu) 
\label{pres}
\end{eqnarray}
Here the inverse gauge boson propagators are given by 
$D_T^{-1}=q^2-q_0^2+\Pi_T+\Pi_{\rm vac}$,  $D_L^{-1}=q^2-q_0^2+\Pi_L+\Pi_{\rm vac}$, and
$D_{\rm vac}^{-1}=q^2-q_0^2+\Pi_{\rm vac}$,
where $\Pi_{T,L,vac}$ are the matter and vacuum contributions to the
gauge boson self-energy produced by an undressed one-loop fermion diagram.
This expression becomes exact in the limit of large flavor number $N_f$
and has been studied in Refs.~\cite{Moore:2002md,%Ipp:2003zr,
Ipp:2003jy} 
and used to
test the behavior of perturbation theory at finite temperature and
chemical potential. At finite $N_f$, Eq.~(\ref{pres}) with
$\Pi$ including also the leading contributions from gluon loops
still collects
all infrared-sensitive contributions up to and including three-loop order
\cite{Vuorinen:2003fs}. We shall however find that
all contributions from gluon loops to $\Pi$ enter the specific heat
only at and beyond order $\g^2 T^3$, and thus will be negligible
%in the regime $T\ll g\mu$.
for $T\ll\mu$ compared to the terms we shall keep.

In the following we will always drop temperature independent 
terms in the pressure, since they do not
contribute to the specific heat at low temperature.
%To the accuracy of the low-temperature expansion that we shall
%achieve we need the low-temperature limit of the transverse
%and longitudinal propagator expanded in $q_0$ as follows

%%...explicit form of the self energy []

%\begin{eqnarray}
%\re \,D_{T}^{-1} & = & q^{2}\left(1+O(\g^2)\right) \label{approxReDT} \nonumber\\
% &+&\!\! \left( \frac{\g ^{2}\mu ^{2}}{\pi ^{2}q^{2}}-1+O(\g^2 q^0)+
%O(\g^2 q^2/\mu^2)\right) q_{0}^{2}\nonumber \\
% &+&\!\! O(\g^2 q_{0}^{4}),
%\end{eqnarray}
%\be
%\im \,D_{T}^{-1} = -\frac{\g ^{2}q_{0}}{48\pi q^{3}}\left( q^{2}-q_{0}^{2}\right) \left( 12\mu ^{2}+3q^{2}+q_{0}^{2}\right) \theta (2\mu -q)\label{approxImDT} 
%\ee
%where $\g^2=e^2N_f$ for QED and $g^2N_f/2$ for QCD.

\subsection{Transverse contribution}\label{subsectT}
\begin{figure}
{\centering \includegraphics[scale=.9]{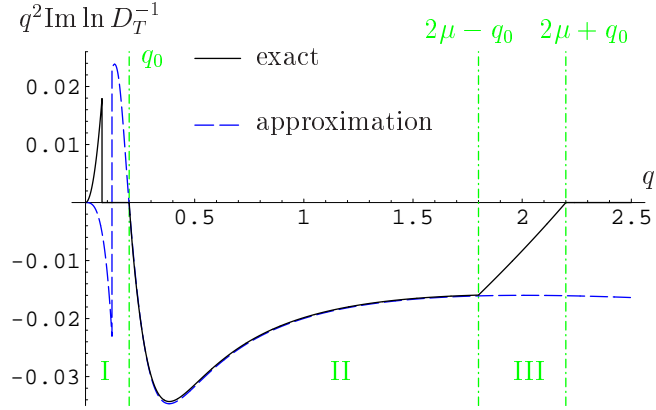} \par}

\caption{Integrand for the \protect$ q\protect $-integration \protect$ q^{2}\im \ln (q^{2}-q_{0}^{2}+\Pi _{T}+\Pi _{\vac })\protect $
for \protect$ \mu =\muMS /2=1\protect $, \protect$ q_{0}=0.2\protect $,
\protect$ \g ^{2}=1\protect $. The solid line shows the exact
result that follows from the full one-loop self
energy expressions at \protect$ T=0\protect $, the dashed line shows the result with the approximations of Eqs.~(\ref{pitapproxr}) and (\ref{h56}).
The parameter \protect$ q_{0}=0.2\protect $
is chosen this large as to clearly show the three different ranges.
As discussed in the text, 
%the error introduced by changing regions I and III
%by our approximation is of order \protect$ O(q_{0}^{3})\protect $
%or higher, whereas 
the main contribution only comes
from region II.
%when the dashed line is integrated from \protect$ q_{0}\protect $
%to \protect$ 2\mu \protect $. 
\label{figintegrand}}
\end{figure}

The $n_b$-part of the contribution of the transverse gluons to the pressure is given by
\begin{equation}
  {P_T\over N_g}=-2\int {d^3q\over(2\pi^3)}\int_0^\infty{dq_0\over\pi}\,n_b\,
  \imag\ln(q^2-q_0^2+\Pi_T+\Pi_{\rm vac}). \label{cvh1}
\end{equation}
%Through order $T^4\ln T$ 
As long as $T\ll\mu$,
it is sufficient to take the self energy at zero temperature,
which is given by a single fermion loop.
From the explicit form of the self energy \cite{Ipp:2004qt} we see that
the $q$-integration naturally
splits into three regions:
$q<q_0$ (I), $q_0<q<2\mu-q_0$ (II), and $2\mu-q_0<q<2\mu+q_0$ (III),
see Fig.\ \ref{figintegrand}.

%In the region I we have the estimate %\cite{Andi}
%\begin{equation}
%  {|P_T^{(1)}|\over N_g}<{1\over\pi^3}\int_0^\infty dq_0\int_0^{q_0} dq\,q^2 n_b{\pi\over2}=
%  {\pi^2T^4\over90},
%\end{equation}
%which means that this contribution is negligible compared to the terms which we will find below. 

Usually
region I contains the ideal-gas pressure of the gluons, $\pi^2 T^4/45$
per gluon,
and perturbative corrections $\propto \g^2 \mu^2 T^2$. However, at
low temperatures $T\ll \omega_p\sim \g\mu$, these contributions
are suppressed by a factor $e^{-\omega_p/T}$ which goes to
zero with all derivatives vanishing and thus do not contribute
to the low-temperature series. All other contributions
from region I are suppressed by further powers of $\g$.

In region III we may expand the self energy about $q=2\mu$ and $q_0=0$. Then one finds that
this contribution is of higher order in $\g$.
We conclude that we may restrict our attention to the region II.

In region II the Bose-Einstein factor and the leading term in the gluon self energy
set the characteristic scales.
Since $q_0\sim T \ll \g\mu$, this is dominated by the
well-known \cite{Weldon:1982aq}
dynamical screening pole at imaginary
$q$ with $|q|\simeq (\g^2\mu^2 q_0/(4\pi))^{1/3}$ and we have
\begin{equation}\label{dynscrscale}
  q_0\sim T,\qquad q\sim(\g^2\mu^2 T)^{1/3}
\end{equation}
in the infrared-sensitive part of region II.

We shall therefore
perform an expansion with respect to a parameter $b$ defined by
\begin{equation}
  b:=\left({T\over \g\mu}\right)^{1/3}.
\end{equation}
It turns out that the following approximation of the gluon self energy is sufficient through order
$T^3\ln T$ in the entropy (see Fig.\ \ref{figintegrand}), 
\begin{eqnarray}\label{pitapproxr}
  &&\real\Pi_T(q_0,q)\simeq{\g^2\over\pi^2}\left({\mu^2q_0^2\over q^2}-{\mu^2q_0^4\over3q^4}\right),\\
  &&\imag\Pi_T(q_0,q)\simeq{\g^2\over4\pi}\left(-{\mu^2q_0\over q}+{\mu^2q_0^3\over q^3}-{q q_0\over4}\right). 
  \label{h56}
\end{eqnarray}
The first two terms in both lines are the leading terms of an expansion of the HDL self energy 
\cite{Braaten:1990mz,Altherr:1992mf,Vija:1995is,Manuel:1996td}, 
in powers of $q_0$. Naively counting powers of $b$ in the integrand one would conclude that only these terms are 
responsible for the terms of order
$b^6$ through $b^{12}$ in the pressure (perhaps with additional factors of $\ln b$). 
In principle this is correct, but one should keep in mind that the
integration limits of the $q$-integration depend on $\mu$ and $q_0$, which might
invalidate a naive power counting. However, it turns out that there is only one instance
where a term which is formally  suppressed in the naive power counting scheme has to be included in
the self energy (see \cite{AGDiss}
for a rigorous proof):
It is the term of order $b^6$ in the pressure, 
where one also finds a contribution from the last term in Eq. (\ref{h56}),
which is beyond the HDL approximation (but it is still included in the large-$N_f$ limit.) 
The $b^6$ term plays a special role anyway,
as this is the only term where we also get 
a contribution from the non-$n_b$-part (see below).

Introducing dimensionless integration variables %\footnote{This substitution also shows
%the characteristic scales: $q_0\sim T$, $q\sim (\g^2\mu^2T)^{1/3}\gg T$.}
$x$ and $y$ via $q_0=b^3\g\mu y$ and $q=b \g\mu (x/(4\pi))^{1/3}$,
we find after expanding the integrand with respect to $b$,
\begin{eqnarray}
  &&{P_T^{\rm II}\over N_g}\simeq {\g^4\mu^4\over 12\pi^4}\int_0^\infty 
  dy{1\over e^y-1}\int_{x_{min}}^{x_{max}}dx
  \bigg[b^6\arctan\left({y\over x}\right)\nonumber\\
  &&+{y(\g^2x^2-64y^2)\over8 (2\pi^2x)^{1/3}(x^2+y^2)}b^8+{32(2x)^{1/3}y^5\over\pi^{4/3}(x^2+y^2)^2}b^{10}
  \nonumber\\
  &&-{32y^5(24x^2y^2-8y^4+\pi^2(x^2+y^2)^2)\over3\pi^2x(x^2+y^2)^3}b^{12}+{O}(b^{14})\bigg],
  \label{cvh4}
\end{eqnarray}
with $x_{min}=4\pi b^6y^3$ and $x_{max}=4\pi(2-b^3\g y)^3/(b\g)^3$.
In the coefficients of this expansion we have written down only those terms which do not
ultimately lead to terms that are suppressed by explicit positive powers of $\g$.
The integrations are now straightforward, and we find
%\begin{eqnarray}
%  &&\!\!\!\!\!\!\!\!\!{P_T^{\rm II}\over N_g}\simeq \g^4\mu^4
%  \bigg[{b^6\over72\pi^2}\left(\ln\left({32\pi\over(b\g)^3}\right)+\gamma_E
%  -{6\over\pi^2}\zeta^\prime(2)+{3\over2}\right)
%  -{2^{2/3}\Gamma\left({8\over3}\right)\zeta\left({8\over3}\right)\over3\sqrt{3}\pi^{11/3}}b^8
%  \nonumber\\
%  &&\qquad+{8\,2^{1/3}\Gamma\left({10\over3}\right)\zeta\left({10\over3}\right)
%  \over9\sqrt{3}\pi^{13/3}}b^{10}
%  +{16(\pi^2-8)\over45\pi^2}b^{12}\ln b+{O}(b^{12})+{\cal R}\bigg], \label{cvh5}
%\end{eqnarray}
%where ${\cal R}$ denotes the contribution from the ${O}(b^{14})$-terms in Eq. (\ref{cvh4}),
%which can be shown to be negligible through order $b^{12}\ln b$. Actually they start to contribute 
%already at the order $b^{12}$ (without $\ln$) in the pressure. This indicates that the naive power counting 
%breaks down at this order, which is related to the $q_0$-dependence of the lower limit of the 
%$q$-integration. 
\begin{eqnarray}
  &&\!\!\!\!\!\!\!\!\!{P_T^{\rm II}\over N_g}\simeq \g^4\mu^4
  \bigg[{b^6\over72\pi^2}\left(\ln\left({32\pi\over(b\g)^3}\right)+\gamma_E
  -{6\over\pi^2}\zeta^\prime(2)+{3\over2}\right)
  -{2^{2/3}\Gamma\left({8\over3}\right)\zeta\left({8\over3}\right)\over3\sqrt{3}\pi^{11/3}}b^8
  \nonumber\\
  &&\qquad+{8\,2^{1/3}\Gamma\left({10\over3}\right)\zeta\left({10\over3}\right)
  \over9\sqrt{3}\pi^{13/3}}b^{10}
  +{16(\pi^2-8)\over45\pi^2}b^{12}\ln b+\tilde c_T b^{12}+{O}(b^{14})\bigg]. \label{cvh5}
\end{eqnarray}
The evaluation of the constant $\tilde c_T$ is a bit more involved 
because one has to
sum up an infinite series of
contributions from the infrared region. 
This calculation is performed in the Appendix and leads to an
integral representation of $\tilde c_T$ that we have been able
to evaluate only numerically, with the result
$\tilde c_T=-0.001786743%05
\ldots$.
 
\subsection{Longitudinal contribution}\label{subsectL}

The $n_b$-part of the contribution of the longitudinal gluons to the pressure is given by
\begin{equation}
  {P_L\over N_g}=-\int {d^3q\over(2\pi^3)}\int_0^\infty{dq_0\over\pi}\,n_b\,
  \imag\ln\left({q^2-q_0^2+\Pi_L+\Pi_{\rm vac}\over q^2-q_0^2}\right). \label{cvh14}
\end{equation}
%First let us take the self energy at zero temperature. 
As in the previous section the $q$-integration
decomposes into three parts.

Again, contributions from region III are suppressed by explicit powers
of $\g$ compared to those of the other regions.

The dominant contribution comes from
region II, $q_0<q<2\mu-q_0$. 
Now the characteristic scales are
\begin{equation}
  q_0\sim T,\qquad q\sim \g\mu,
\end{equation}
because of Debye screening of the longitudinal gluons with
leading-order mass $m_D=\g\mu/\pi$.
In a similar way as in the previous section, the gluon self energy can be approximated as
\begin{eqnarray}\label{repilapprox}
  &&\real\Pi_L(q_0,q)\simeq{\g^2\over\pi^2}\left(\mu^2-{2\mu^2q_0^2\over q^2}-{q^2\over12}\right),\\ \label{impilapprox}
  &&\imag\Pi_L(q_0,q)\simeq{\g^2\over2\pi}\left({\mu^2q_0\over q}-{\mu^2q_0^3\over q^3}-{q q_0\over4}\right). 
\end{eqnarray}

We introduce dimensionless
integration variables $y$ and $z$ via $q_0=b^3\g\mu y$, $q=\g\mu z/\pi$. Then we find after expanding
the integrand with respect to $b$,
\begin{eqnarray}
  &&\!\!\!\!\!\!\!\!\!\!\!\!\!\!\!\!\!{P_L^{\rm II}\over N_g}\simeq {\g^4\mu^4\over 16\pi^2}\int_0^\infty 
  dy{1\over e^y-1}\int_{z_{min}}^{z_{max}}dz \nonumber\\
  &&\!\!\!\!\!\!\!\!\!\!\!\!\!\!\!\!\!\times\bigg[b^6{yz(-4\pi^2(1+z^2)+\g^2z^4)\over\pi^4(1+z^2)^2}
  +b^{12}{y^3(\pi^2-12(z^2+1))\over3z(1+z^2)^3}+{O}(b^{18})\bigg],\quad \label{cvh9}
\end{eqnarray}
with $z_{min}=b^3y\pi$ and $z_{max}=(2/\g-b^3y)\pi$.
In the coefficients of this expansion we have written down only those terms which do not
lead ultimately 
to terms that are suppressed by explicit positive powers of $\g$.
The integrations are now straightforward, and we find
%\begin{equation}
%  {P_L^{\rm II}\over N_g}\simeq \g^4\mu^4\left[{b^6\over48\pi^2}\left(1+\ln\left({\g^2\over4\pi^2}\right)\right)
%  +{\pi^2(12-\pi^2)\over240}b^{12}\ln b+{\cal R}^\prime\right], \label{cvh10}
%\end{equation}
%where ${\cal R}^\prime$ denotes the contribution from the ${O}(b^{18})$-terms in Eq. (\ref{cvh9}),
%which can again be shown to be negligible through order $b^{12}\ln b$.
\begin{equation}
  {P_L^{\rm II}\over N_g}\simeq \g^4\mu^4\left[{b^6\over48\pi^2}\left(1+\ln\left({\g^2
\over4\pi^2}\right)\right)
  +{\pi^2(12-\pi^2)\over240}b^{12}\ln b+\tilde c_L^{\rm II}b^{12}+{O}(b^{18}\ln b)\right]. \label{cvh10}
\end{equation}
The constant $\tilde c_L^{\rm II}$ can be determined by summing up IR enhanced contributions in analogy to the constant $\tilde c_T$ of Sect.~\ref{subsectT},
and its integral representation is given in Eq.~(\ref{ctildeL}).
Numerically, we get $\tilde c_L^{\rm II}\approx0.11902569216\ldots$.
%One finds 
%\begin{eqnarray}
%  &&\tilde c_L^{\rm II}={1\over8640}\Big(3\pi^4-2\pi^2(12-\pi^2)(-17+6\gamma_E-6\ln(\pi)-540{\zeta^\prime(4)\over\pi^4}
%  \Big)\nonumber\\
%  &&\qquad-{\pi\over30}\int_1^\infty dz\Bigg({\pi(\pi^2+12(-1-z^2+z^3))\over24z}\nonumber\\
%  &&\qquad+z^2\arctan\bigg[-{2z\over\pi}+{1\over\pi}\ln\left({z+1\over z-1}\right)\bigg]\Bigg)\nonumber\\
%  &&\quad\simeq 0.11902569216\ldots
%\end{eqnarray}     

In contrast to the case of transverse polarizations, there is now however
also a contribution from region I, of the order of $\g^4 \mu^4 b^{12} =
T^4$. The term involving 
$\imag\ln(q^2-q_0^2+\Pi_L+\Pi_{\rm vac})$ is again exponentially suppressed
for $T\ll \omega_p\sim \g\mu$. However, at these temperatures
the term involving
$\imag\ln(q^2-q_0^2)$ contributes the equivalent of an ideal-gas
pressure of one bosonic degree of freedom, but with negative sign,
leading to
\be
{P_L^{\rm I}\over N_g}\simeq -{\pi^2 T^4\090} + O(e^{-\omega_p/T})
\ee
so that $\tilde c_L=\tilde c_L^{\rm II}-\pi^2 T^4/90=0.009363421\ldots$.

\subsection{Non-$n_b$ contribution}\label{subsectnon}

The nonanalytic terms in the low-temperature expansion of Eq. (\ref{pres})
all come from the parts of the integrals involving the Bose distribution
$n_b$.
The non-$n_b$ parts in Eq. (\ref{pres}) are less IR singular and
can be calculated by expanding out the self energy diagrams.
We can determine their contribution by the observation that
at two-loop order also the $n_b$ part is IR safe and given by
%Expanding again with respect to $b$, 
%we find only one contribution from this part which is not suppressed by powers of $\g$, namely a
%term of order $\g^2\mu^2T^2$. This is a two-loop contribution, which can be calculated most easily by
%considering first the two-loop $n_b$-part,
\begin{equation}
  {P_{n_b}^{\rm 2-loop}\over N_g}=-\int {d^3q\over(2\pi)^3}\int_0^\infty {dq_0\over\pi}n_b\,\imag 
  \left({2\Pi_T+\Pi_L\over q^2-q_0^2-i\epsilon}\right).
\end{equation}
In this integral we have two contributions, one from the real parts and one from the imaginary parts of the 
gluon self energies. 
One finds that these two contributions cancel precisely at the order $\g^2\mu^2T^2$. 
(As above one finds that the $T=0$ gluon self energies are sufficient at this order.)
Therefore the two-loop non-$n_b$
contribution has to be equal to the standard perturbative result at order $\g^2\mu^2T^2$ \cite{Kap:FTFT},
\begin{equation}
%  {P_{{\rm non-}n_b}\over N_g}
{1\0N_g}\biggl[P_{{\rm non-}n_b}-P_{{\rm non-}n_b}\Big|_{T=0}\biggr]
=-{\g^2\mu^2T^2\over16\pi^2}+\mathcal{O}(\g^2T^4).
\end{equation}
%This result is consistent with the numerical calculations \ldots

\subsection{Combined result}

Our final expression for the leading temperature-dependent contribution to
the interaction pressure in the regime $T\ll g\mu$ is contained in
\begin{equation}
  \Delta P=P-P^0=P_T^{\rm II}+
P_L^{\rm I}+
P_L^{\rm II}+P_{{\rm non-}n_b}-P^0+{O}(g^4\mu^4),
\end{equation}
where $P^0$ is the ideal-gas pressure,
%where \cite{Ipp:2003cj}
%\begin{equation}
%  P_{{\rm non-}n_b}\simeq-{\g^2N_g\mu^2 T^2\over16\pi^2}.
%\end{equation}
and explicitly reads
\bea\label{finalpressure}
{1\0N_g}\biggl[\Delta P-\Delta P\Big|_{T=0}\biggr]&=&
  {\g^2\mu^2T^2\over72\pi^2}\left(\ln\left({4\g\mu\over\pi^2T}\right)
  +\gamma_E-{6\over\pi^2}\zeta^\prime(2)-{3\02}\right)\nonumber\\
  &&-{2^{2/3}\Gamma\left({8\over3}\right)\zeta\left({8\over3}\right)\over3\sqrt{3}\pi^{11/3}}
  T^{8/3}(\g\mu)^{4/3}
  +8{2^{1/3}\Gamma\left({10\over3}\right)\zeta\left({10\over3}\right)
  \over9\sqrt{3}\pi^{13/3}}T^{10/3}(\g\mu)^{2/3}\nonumber\\
  &&+{2048-256\pi^2-36\pi^4+3\pi^6\over2160\pi^2}T^4
  \left[\ln\left({\g\mu\over T}\right)+\bar c\, \right]
\nonumber\\&&+{O}(T^{14/3}/(\g\mu)^{2/3})
+{O}(g^4\mu^2 T^2 \ln T).
\eea
%We have verified the correctness of the coefficients appearing
%in this low-temperature expansion also by evaluating numerically
%the relevant integrals appearing in $P_T$ and $P_L$ 
%and doing so
%we have in fact been able to extract the coefficient under the
%log of the $T^4 \ln T$ term 
%%(which also requires
%%keeping the contributions from region III),  %g^2 suppressed
%with the result
%\be
%\tilde c= -2.26969614669423\ldots
%\ee
where the constant $\bar c$ is given by
\begin{eqnarray}
  \bar c&=&\gamma_E-90{\zeta^\prime(4)\over\pi^4}-{31\over12}+{1\over 2048-256\pi^2-36\pi^4+3\pi^6}\times
  \nonumber\\
  &&\quad\times\Bigg\{
  3\pi^4(12-\pi^2)\ln\pi+128(\pi^2-8)\ln(4\pi)+3\pi^2(29\pi^2+32)\nonumber\\
  &&\qquad-72\pi^3\int_1^\infty dz\Bigg[{1024+\pi^6-64\pi^2(2+3z^2)+12\pi^4(-1-z^2+3z^3)\over24\pi^3z}
  \nonumber\\
  &&\qquad\quad+2z^2 \arctan\bigg[{\pi(1-z^2)\over 2z+(z^2-1)\ln\left({z+1\over z-1}\right)}\bigg] %\nonumber\\
%  &&\qquad
+z^2\arctan\bigg[-{2z\over\pi}+{1\over\pi}\ln\left({z+1\over z-1}\right)\bigg]\Bigg]\Bigg\}\nonumber\\
&\approx&  %-2.2696961466942302
4.099348512039
\ldots.
\end{eqnarray}
The terms involving logarithms and fractional powers of $T$ all
come from the cut contribution of region II, whereas $P^{\rm I}-P^0=
-{3\02}P^0+O(e^{-\omega/T})$ only contributes to $\bar c$.

From Eq.~(\ref{finalpressure}) one can obtain the 
entropy density through $\cS=\left({\partial P/\partial T}\right)_\mu$
and the specific heat through \cite{LL:V-Cv}
%Inserting the explicit expression for $P$ into
\begin{equation}\label{CvLL}
%  \cS=\left({\partial P\over\partial T}\right)_\mu,\quad 
  \mathcal C_v=T\left(\left({\partial\cS\over\partial T}\right)_\mu-
  \left({\partial\cN\over\partial T}\right)_\mu^2\left({\partial\cN\over\partial\mu}\right)_T^{-1}\right),
\end{equation}
with $\cN=\left({\partial P/\partial \mu}\right)_T$, which in
the ideal-gas limit reads
\be
\mathcal C_v^0=N N_f \left[ {\mu^2 T\03} + T^3
\left ({7\pi^2\015}-{4\mu^2\03 T^2+9\mu^2/\pi^2} \right) \right]
+N_g {4\pi^2 T^3\015} .
\ee

For the interaction part of the specific heat only the logarithmic
derivative of the entropy in formula (\ref{CvLL}) contributes,
and is given explicitly by
\begin{eqnarray}\label{finalcv}
  &&\!\!\!\!\!\!\!\!\!\!\!\!
  {\mathcal C_v-\mathcal C_v^0\over N_g}={\g^2\mu^2 T\over36\pi^2}\left(\ln\left({4\g\mu\over\pi^2T}\right)+\gamma_E
  -{6\over\pi^2}\zeta^\prime(2)-3\right)\nonumber\\
  &&-40{2^{2/3}\Gamma\left({8\over3}\right)\zeta\left({8\over3}\right)\over27\sqrt{3}\pi^{11/3}}
  T^{5/3}(\g\mu)^{4/3}
  +560{2^{1/3}\Gamma\left({10\over3}\right)\zeta\left({10\over3}\right)
  \over81\sqrt{3}\pi^{13/3}}T^{7/3}(\g\mu)^{2/3}\nonumber\\
  &&+{2048-256\pi^2-36\pi^4+3\pi^6\over180\pi^2}T^3
  \left[\ln\left({\g\mu\over T}\right)+\bar c-{7\012}\right]\nonumber\\
&&+{O}(T^{11/3}/(\g\mu)^{2/3})
+{O}(g^4\mu^2 T \ln T).  \label{cv1}
\end{eqnarray}
%where the constant $M$ is still undetermined.
We remark that in Eq. (\ref{cv1}) one could replace $\mu$ with $\mu(T)$, as given in Eq. (\ref{x34}),
as this would modify the result only beyond the terms of order $T^3$
since $T/(g\mu)\ll 1$.

Using the method described in Sect.~\ref{subsectT} -- \ref{subsectnon}, one can in principle compute the coefficients of
higher terms in the expansion of $\mathcal C_v$ with respect to $b$. This is straightforward for the
coefficients of the fractional powers and the logarithmic terms, where one only has to include
higher orders in the expansion of the HDL gluon self energy with respect to $q_0$, see Eqs.~(\ref{pitapproxr}), (\ref{h56}), (\ref{repilapprox}), 
(\ref{impilapprox}).   
For the terms of order $T^5$, $T^7$, $T^9$ etc., however, one has to sum up again IR enhanced terms,
in a similar way as in the calculation of $\tilde c$ 
described in the Appendix. 

The low-temperature expansion that we have carried out has assumed
that $T\ll \g\mu$ as well as $\g\ll1$. 
If we set $T/\mu\sim \g^{1+\delta}$ with $\delta>0$, we
find that
the terms in the expansion (\ref{finalcv}) correspond to
the orders $\g^{3+\delta}\ln(c/\g)$, $\g^{3+(5/3)\delta}$,
$\g^{3+(7/3)\delta}$, and $\g^{3+3 \delta}\ln(c/\g)$, respectively, 
with a truncation error of
the order $\g^{3+(11/3)\delta}$ from higher terms in the ring diagrams.
We have neglected perturbative corrections to these terms, which
at a minimum arise at the order $\g^{5+\delta}\ln(c/\g)$.

One might suspect that higher order terms could involve
also higher powers of $\g^2 \ln T$, which could resum into
a leading term $\mu^2 T^{1+O(\g^2)}$. However, it has
been argued in \cite{Chakravarty:1995} that the leading
$\g^2 T \ln T^{-1}$ is not modified by higher order corrections in QED,
and this has been corroborated recently by the authors
of Ref.~\cite{Schafer:2004zf} using a high-density effective field
theory also applicable to QCD.
It can therefore be expected that the leading term in the
above low-temperature series remains valid even when the
temperature is so low that $\g^2 \ln (\g \mu/T) \gg 1$.

On the other hand, the higher terms of the low-temperature
expansion involving fractional powers 
$T^{(2n+3)/3}$ with $n\ge1$
remain more important than the undetermined perturbative 
corrections (which are suppressed by explicit powers of $\g^2$) only
when $\delta < 3/n$.

\section{HDL resummation}
\label{sect4plus}

As we have seen in Sect.~\ref{sectIV}, the nonanalytic terms in
the low-temperature expansion of the thermodynamic potential
are determined by HDL contributions to the gluon self-energy.
Terms beyond the HDL approximation are relevant for contributions
from hard momenta $q\sim\mu$, yielding a term of order $\g^2\mu^2 T^2$
in the temperature-dependent part of the pressure. However, this is
a perturbative piece that can be identified as a
two-loop contribution without the need for resummations. 
When this contribution is subtracted from the full one-loop expression, the
remainder is dominated by soft momenta $q\ll\mu$ and the HDL
approximation is sufficient. 

In this section we shall consider the full HDL-resummed
ring diagrams, which allows us to relax the requirement $T\ll \g\mu$,
under which the above low-temperature series is meaningful, to
only $T\ll \mu$. When expanded around $T=0$, the one-loop
HDL-resummed thermodynamic potential contains all the
anomalous terms of the low-temperature series (\ref{finalpressure}).
However, as we have already seen
there are also terms from region I which behave as $\sim e^{-\omega_p/T}$
and thus do not show up at any finite order of the low-temperature
series. Nevertheless, such terms become important for $T\sim\g\mu$.
By numerically evaluating the full HDL-resummed expression we can
capture their effect as well and study the behavior of entropy
and specific heat for all temperatures $T\ll \mu$.

\subsection{Separation of hard and soft contributions}

In the transverse sector,
the one contribution in Eq.~(\ref{cvh4}) 
from a non-HDL term in the gluon self-energy
can also be written as
\bea
&&{1\0N_g}\biggl[P^{{\rm II,non-HDL}}_T-P^{{\rm II,non-HDL}}_T\Big|_{T=0}\biggr]\nonumber\\&&=
-{1\0\pi^3}\int_0^\infty dq_0\, n_b(q_0)
\int_{q_0}^{2\mu} dq\,q^2\,{\imag\Pi_T^{(2)}\0q^2-q_0^2} \simeq {\g^2\mu^2 T^2
\048\pi^2}
\eea
where $\Pi_T^{(2)}\equiv\Pi_T-\Pi_T^{\rm HDL}$ with \cite{Ipp:2004qt}
\be
\imag\Pi_T^{(2)}\simeq-{\g^2 q q_0\016\pi},\qquad q\gg q_0,
\ee
while $\imag\Pi_T^{(2)}\to0$ for $q\to q_0$.

Similarly, in the longitudinal sector the non-HDL contribution
to Eq.~(\ref{cvh9}) is
\bea
&&{1\0N_g}\biggl[P^{{\rm II,non-HDL}}_L-P^{{\rm II,non-HDL}}_L\Big|_{T=0}\biggr]
\nonumber\\&&=
-{1\02\pi^3}\int_0^\infty dq_0\, n_b(q_0)
\int_{q_0}^{2\mu} dq\,q^2\,{\imag\Pi_L^{(2)}\0q^2-q_0^2} \simeq {\g^2\mu^2 T^2
\048\pi^2}
\eea
where 
\be
\imag\Pi_L^{(2)}\simeq-{\g^2 q q_0\08\pi},\qquad q\gg q_0,
\ee
and again $\imag\Pi_L^{(2)}\to0$ for $q\to q_0$.

The HDL part of the gluon self-energy, explicitly given by
\begin{eqnarray}
\Pi^{\rm HDL}_{T}(q_{0},q) &=& m_D^2\frac{q_{0}^{2}}{2q^{2}}\left( 1+\frac{q_{0}^{2}-q^{2}}{2qq_{0}}\log \frac{q_{0}-q}{q_{0}+q}\right) ,\label{PiTHDL}\\
\Pi^{\rm HDL}_{L}(q_{0},q) &=& m_D^2\frac{q^{2}-q_{0}^{2}}{q^{2}}\left( 1+\frac{q_{0}}{2q}\log \frac{q_{0}-q}{q_{0}+q}\right), 
\label{PiLHDL}
\end{eqnarray}
with $m_D=\g\mu/\pi$
is the leading-order contribution at small $q_0,q\ll T$ provided $T\ll\mu$.
In order to retain all contributions that are nonanalytic in $T$ at $T=0$,
the HDL self-energies 
need to be kept unexpanded in
\bea\label{PHDL}
&&{1\0N_g}\biggl[P^{{\rm HDL}}-P^{{\rm HDL}}\Big|_{T=0}\biggr]\nonumber\\&=&
-{1\02\pi^3}\int_0^\infty dq_0\, n_b(q_0)
\int_0^{2\mu} dq\,q^2\,\biggl[ 2\,\im \ln \left( q^{2}-q_{0}^{2}+\Pi^{\rm HDL} _{T}\right)
+\im \ln \left( \frac{q^{2}-q_{0}^{2}+\Pi^{\rm HDL} _{L}}{q^{2}-q_{0}^{2}}\right)
\biggr]\quad
\eea
where we have dropped contributions from region III as being suppressed
by explicit powers of $\g$.

Individually, the transverse and the longitudinal contributions
depend logarithmically on the upper integration boundary, because
\bea
&&\im \ln \left( q^{2}-q_{0}^{2}+\Pi^{\rm HDL} _{T}\right) \to -\frac{\g ^{2}\mu ^{2}}{4\pi }\frac{q_{0}}{q^{3}}\\
&&
\im \ln \left( \frac{q^{2}-q_{0}^{2}+\Pi^{\rm HDL} _{L}}{q^{2}-q_{0}^{2}}\right)
\to +\frac{\g ^{2}\mu ^{2}}{2\pi }\frac{q_{0}}{q^{3}}
\eea
at large $q$. However, the combined expression (\ref{PHDL}) is saturated
by soft momenta $q\ll\mu$, and
the upper integration limit $2\mu$ can be sent to infinity. This
just amounts to dropping terms that are suppressed by explicit powers
of $\g$.

The only other contribution that needs to be taken into account
is $P_{{\rm non-}n_b}$, which as discussed in
Sect.~\ref{subsectnon} can be treated perturbatively
to the order under consideration. Put together, the final result
for $\Delta P=P-P^0$ is
\bea\label{Pfull}
{1\0N_g}\biggl[\Delta P-\Delta P\Big|_{T=0}\biggr]&=&-{\g^2\mu^2T^2\048\pi^2}
-{1\02\pi^3}\int_0^\infty dq_0\, n_b(q_0)
\int_0^\infty dq\,q^2\,\biggl[ 2\,\im \ln \left( q^{2}-q_{0}^{2}+\Pi^{\rm HDL} _{T} \0 q^{2}-q_{0}^{2} \right )\nonumber\\&&\qquad
+\im \ln \left( \frac{q^{2}-q_{0}^{2}+\Pi^{\rm HDL} _{L}}{q^{2}-q_{0}^{2}}\right)
\biggr] + O(g^4\mu^2T^2),
\eea
where we have subtracted the ideal-gas pressure of the gluons
by including a free propagator in the argument of the first logarithm.

As indicated, this expression provides the leading terms in the
temperature dependent part of the pressure and therefore the
leading terms in entropy and specific heat. For $T\ll \g\mu$ 
the contribution to the pressure (as opposed to entropy and specific heat)
is subleading compared to the three-loop
result for the zero-temperature pressure obtained by
Freedman and McLerran \cite{Freedman:1977dm}, 
Baluni \cite{Baluni:1978ms}, and Vuorinen \cite{Vuorinen:2003fs}.
However, when $T \gtrsim \g\mu$, its
magnitude is comparable to $\g^4\mu^4\ln(c/g)$, the highest known term in the 
perturbative result of the $T=0$ pressure,
and thus provides an extension of the
result of Refs.~\cite{Freedman:1977dm,Baluni:1978ms,Vuorinen:2003fs} to non-zero temperatures in
the domain $T\ll\mu$.

\subsection{HDL quasiparticle pole contribution}

In region I, i.e.\ $0\le q\le q_0$, the HDL propagator has single
poles for $q_0\ge \omega_p=m_D/\sqrt3$ at real $q_0=\omega_{T,L}(q)$,
which allows us to carry out the $q_0$ integration, yielding
\bea\label{PHDLI}
&&{1\0N_g}\biggl[P^{{\rm I,HDL}}-P^{{\rm I,HDL}}\Big|_{T=0}\biggr] %^{pole}
\nonumber\\&=&
-{T\02\pi^2}\int_0^\infty dq\,q^2
\biggl[ 2 \ln\left(1-e^{-\omega_T(q)/T}\right)+
\ln\left(1-e^{-\omega_L(q)/T} \0 1-e^{-q/T}\right)
\biggr]\,.
\eea

At small temperatures $T\ll\omega_p$, the contribution from
transverse polarizations is suppressed by a factor $e^{-\omega_p/T}$
and thus does not show up at any finite order of a low-temperature expansion
in terms of powers and logarithms of $T$.
Using that for small $q\ll\omega_p$ the dispersion law of
transverse gluons is given by \cite{Weldon:1982aq}
\be
\omega_T(q)=\omega_p\left(1+{3q^2\05\omega_p^2}-O(q^4/\omega_p^4)\right)
\ee
one can calculate the leading term as
\be
{1\0N_g}\biggl[P^{{\rm I,HDL}}_T-P^{{\rm I,HDL}}_T\Big|_{T=0}\biggr]
\simeq {5\012} \sqrt{5\omega_p^3 T^5\03\pi^3}e^{-\omega_p/T},
\quad T\ll\omega_p\;.
\ee

As discussed already in Sect.~\ref{subsectL},
the low-temperature contribution of the longitudinal gluons
involves a contribution $\propto T^3$ as well as exponentially
suppressed terms. Using that
\be
\omega_L(q)=\omega_p\left(1+{3q^2\010\omega_p^2}-O(q^4/\omega_p^4)\right)
\ee
one can show that
\be
{1\0N_g}\biggl[P^{{\rm I,HDL}}_L-P^{{\rm I,HDL}}_L\Big|_{T=0}\biggr]
\simeq -{\pi^2\090}T^4+{5\06} \sqrt{5\omega_p^3 T^5\06\pi^3}e^{-\omega_p/T},
\quad T\ll\omega_p\;.
\ee

At temperatures $\omega_p \ll T \ll \mu$ the
dispersion relation of the longitudinal gluons approaches the light-cone
exponentially, which gives an equally exponentially vanishing
contribution to the pressure. The transverse gluons, on the other
hand, tend to the mass hyperboloid $\omega_T(q)\to \sqrt{q^2+m_\infty^2}$
with $m_\infty^2=m_D^2/2$, yielding
\be
{1\0N_g}\biggl[P^{{\rm I,HDL}}-P^{{\rm I,HDL}}\Big|_{T=0}\biggr]
\simeq 2\left(
{\pi^2\090}T^4
-{m_\infty^2\024}T^2 \right)
={\pi^2T^4\045}-{\g^2\mu^2 T^2\024\pi^2},
\quad \omega_p \ll T \ll \mu.
\ee

\subsection{HDL cut contribution}

For $q>q_0$, the HDL self-energies have an imaginary part $\propto m_D^2
=\g^2\mu^2/\pi^2$ which corresponds to collisionless Landau damping
of hard fermions. 
At low $q_0\sim T\ll \g\mu$, this provides the dynamical screening
of quasistatic magnetic fields which is responsible for
the anomalous terms in the 
low-temperature expansion of the specific heat, Eq.~(\ref{dynscrscale}).

At higher temperatures $\omega_p \ll T \ll \mu$, it is instead
electric Debye screening which gives the dominant contribution
from soft momentum scales
\bea
{1\0N_g}\biggl[P^{{\rm HDL}}-P^{{\rm HDL}}\Big|_{T=0}\biggr]
&\simeq& 2\left(
{\pi^2\090}T^4
-{m_\infty^2\024}T^2 \right)+{m_D^3T\012\pi}\nonumber\\&&
={\pi^2T^4\045}-{\g^2\mu^2 T^2\024\pi^2}+{\g^3\mu^3 T\012\pi^4},
\quad \omega_p \ll T \ll \mu.
\eea
(At still higher temperatures $T\gtrsim\mu$, one eventually
has to replace the value of the HDL Debye mass by the
hard-thermal-loop result \cite{LeB:TFT}). 

The appearance of a contribution $\propto m_D^3$ is traditionally
referred to as plasmon effect.
Our full HDL-resummed result (\ref{Pfull}) gives a unified
description of this longitudinal plasmon effect with the
anomalous (non-Fermi-liquid) contributions from transverse quasistatic fields
which are only weakly screened, and interpolates between
these two different effects. As the temperature is lowered,
the longitudinal plasmon term $m_D^3 T$ which is linear in $T$
gradually disappears and gets replaced by a quadratic term
$\propto m_D^2 T^2 \ln(m_D/\mu)$, cf.\ Eq.~(\ref{cvh10}).
This then combines with the leading
anomalous term $\propto m_D^2 T^2 \ln(T^{1/3}m_D^{2/3}/\mu)$
from the transverse sector, cf.\ Eq.~(\ref{cvh5})
where $m_D$ enters through
dynamical screening, and whose logarithmic
dependence on the hard scale $\mu$ cancels that
of the longitudinal sector.

\section{Numerical results}
\label{sectV}

\subsection{Full HDL result versus low-temperature expansion}

We shall now turn to a numerical evaluation of entropy and
specific heat following from the HDL-resummed pressure (\ref{Pfull}).
The corresponding expression for the
entropy density 
is given explicitly by
\bea\label{Sfull}
{1\0N_g}(\mathcal S-\mathcal S^0)&=&-{\g^2\mu^2T\024\pi^2}
-{1\02\pi^3}\int_0^\infty dq_0\, {\6n_b(q_0)\0\6T}
\int_0^\infty dq\,q^2\,\biggl[ 2\,\im \ln \left( q^{2}-q_{0}^{2}+\Pi^{\rm HDL} _{T} \0 q^{2}-q_{0}^{2} \right )\nonumber\\&&\qquad
+\im \ln \left( \frac{q^{2}-q_{0}^{2}+\Pi^{\rm HDL} _{L}}{q^{2}-q_{0}^{2}}\right)
\biggr] + O(\g^4\mu^2T),
\eea
where $\mathcal S^0$ is the ideal-gas entropy density.
Eq.~(\ref{Sfull}) represents the leading interaction term at weak coupling
for all $T\ll\mu$. It is essentially given by one universal function
of the dimensionless variable $T/(\g\mu)$, which we define through
\be
{8\pi^2\0N_g(\g\mu)^2T}(\mathcal S-\mathcal S^0)=:\mathfrak S\left(\textstyle{T\0\g\mu}\right) + O(\g^2),
%\times{T\0\g\mu}\ln{T\0\g\mu})
\ee
and which we have normalized such that the ordinary perturbative
two-loop result \cite{Kap:FTFT} for the low-temperature
entropy corresponds to $\mathfrak S=-1$.

\begin{figure}[t]
%\vspace{-1cm}
\qquad\qquad\includegraphics[%viewport=72 180 540 550,
width=0.75%
\linewidth]{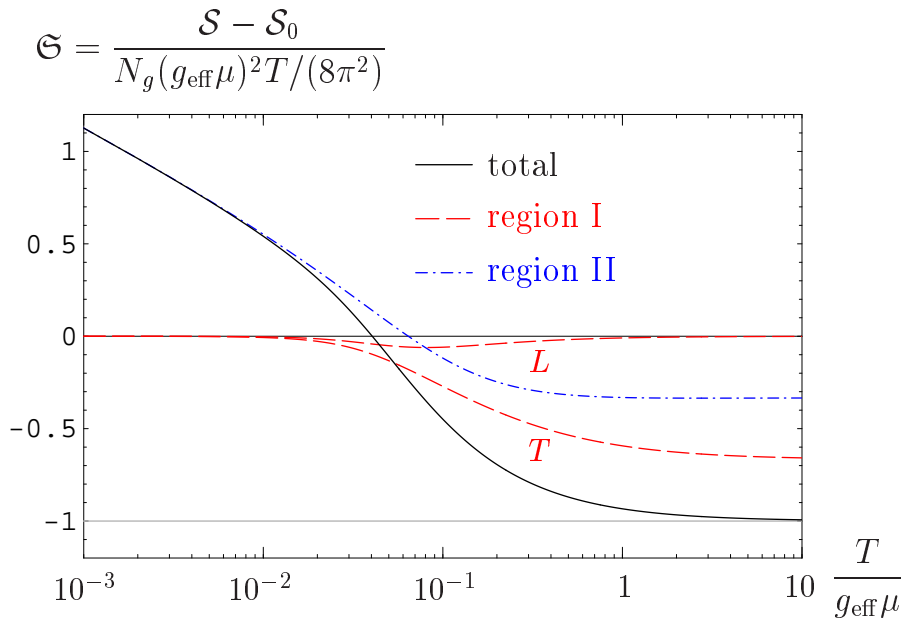}
\caption{The function $\mathfrak S(T/(\g\mu))$
which determines the leading-order interaction contribution
to the low-temperature entropy.
The normalization is such that $\mathfrak S=-1$ corresponds
to the result of ordinary perturbation theory.
The dash-dotted line shows the contribution from
region II, comprising HDL Landau damping and
hard contributions; the two dashed lines
give the transverse (T)
and longitudinal (L) quasiparticle pole 
contributions of region I.
\label{figSdetails}}
\end{figure}

In Fig.~\ref{figSdetails} we display the individual contributions
to $\mathfrak S$ as provided by transverse and longitudinal quasiparticle poles
(region I), and the Landau damping cut (region II).
As one can see, the latter is responsible for the anomalous
behavior of an interaction contribution which is positive
for $T/(g\mu)\le 0.0404\ldots$ and is well reproduced by the
low-temperature series involving logarithms and fractional 
powers of $T/(\g\mu)$. The quasiparticle contributions, on the other
hand, behave as $T^3$ at small temperatures, but cannot be further
expanded about $T=0$ due to terms involving $e^{-\omega_p/T}$.

We have in fact been able to perform the required numerical integrations
with sufficient accuracy to explicitly check all the coefficients
of the low-temperature expansion calculated in Sect.~\ref{sectIV}
(and further ones up to order $T^5 \ln T$ \cite{AGDiss}).

\begin{figure}[t]
%\vspace{-1cm}
\qquad\qquad\includegraphics[width=0.75%
\linewidth]{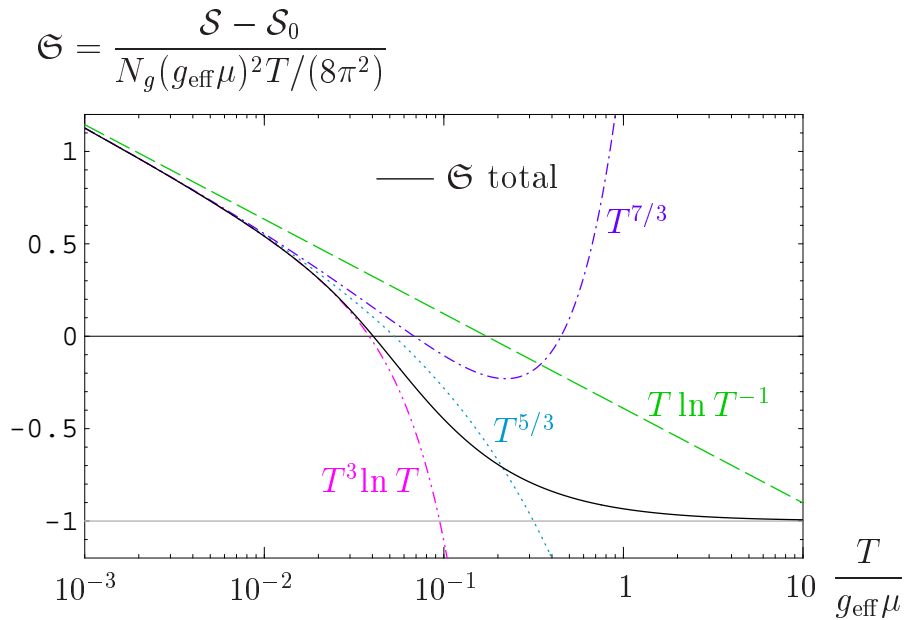}
\caption{The first few orders of the low-temperature series
for the entropy as determined by Eq.~(\ref{finalpressure})
in comparison with the full HDL-resummed result.\label{figconv}}
\end{figure}

In Fig.~\ref{figconv} we compare
the first few orders of the low-temperature series
with the full HDL result. The low-temperature result up to
and including the $T^3\ln T$ contribution to the entropy
is a good approximation for $T/(\g\mu)\lesssim 0.04$
where the anomalous contributions dominate; for larger $T$
the non-expandable $e^{-\omega_p/T}$ terms in the
quasiparticle pole contributions become important numerically.

\subsection{Comparison with nonperturbative large-$N_f$ results}

\begin{figure}[t]
%\vspace{-1cm}
\includegraphics[width=0.75%
\linewidth]{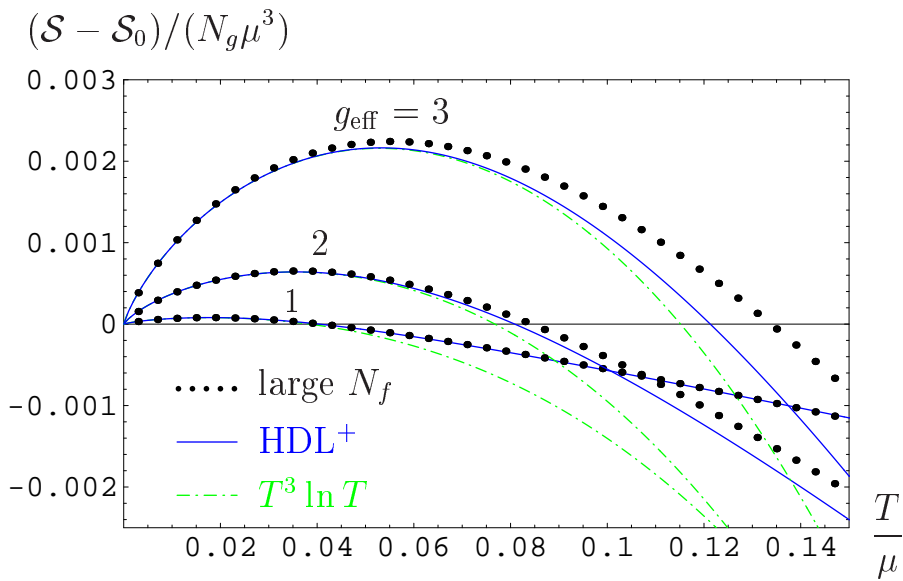}
\caption{Complete entropy density
in the large-$N_f$ limit for the three values
$\g(\bar\mu_{\rm MS}=2\mu)=1,2,3$ (heavy dots), compared with
the full HDL result (solid line) and the
low-temperature series up to and including the 
$T^3 \ln T$ contributions.\label{figSfull}}
\end{figure}

When applying our results to QED as well as QCD,
the range of $T/\mu$ where one finds an excess of entropy and
specific heat over the ideal-gas result will be the larger
the higher the coupling is. 
However, we then have to expect more important
perturbative corrections which are suppressed parametrically by further
powers of $g^2$. In order
to assess their importance, we compare with the special but
exactly solvable case of infinite flavor number.
This is done in Fig.~\ref{figSfull} 
for the three values
$\g(\bar\mu_{\rm MS}=2\mu)=1,2,3$ 
where the heavy dots give the
nonperturbative large-$N_f$ result of Ref.~\cite{Ipp:2003jy} and
the full line represents the full HDL result 
(solid line,
denoted by HDL$^+$ to remind of the inclusion of
hard, perturbative terms). Also given is the
low-temperature series up to and including the 
$T^3 \ln T$ contributions.

%This is done in Fig.~\ref{figST} first only for the transverse
%$n_b$-contributions which are dominated by the anomalous terms.
%We obtain a rather satisfactory agreement with the
%exact large-$N_f$ result in the temperature range considered.
%%corresponding
%%to decreasing terms in the low-temperature expansion.

\begin{figure}
%\vspace{-1cm}
\includegraphics[width=0.75%
\linewidth]{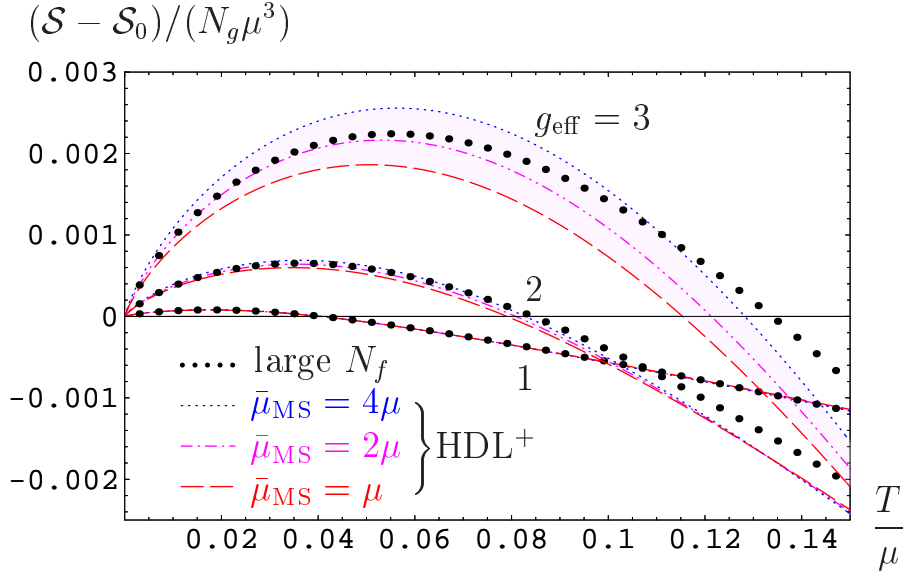}
\caption{Complete entropy density
in the large-$N_f$ limit for the three values
$\g(\bar\mu_{\rm MS}=2\mu)=1,2,3$ (heavy dots), compared with
the HDL-resummed result
when in the latter the renormalization scale
is varied by a factor of 2 around $\bar\mu_{\rm MS}=2\mu$.\label{figSfullscd}}
\end{figure}

\begin{figure}[t]
\qquad\qquad\includegraphics[width=0.75%
\linewidth]{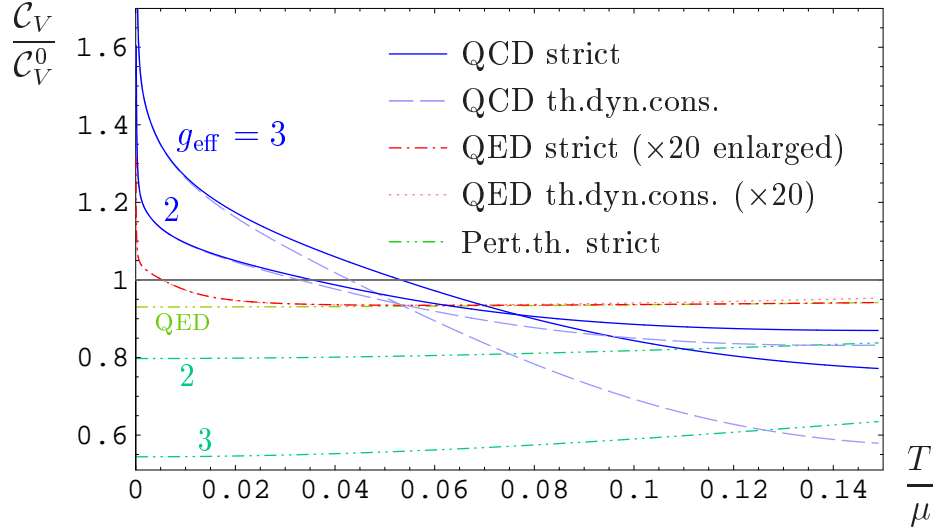}
\caption{The HDL-resummed result for the specific heat
$\mathcal C_v$,
normalized to the ideal-gas value
for $\g=2$ and 3 corresponding
to $\alpha_s\approx 0.32$ and $0.72$ in two-flavor QCD,
and
$\g\approx 0.303$ for QED.
The results labelled ``strict'' does not include anomalous
contributions in the second term of Eq.~(\ref{CvLL}) where
the would be of higher order in $\g$, whereas
``th.dyn.cons.'' refers to a less systematic
but thermodynamically consistent evaluation.
The deviation of the QED result from
the ideal-gas value is enlarged by a factor of 20
to make it more visible.
\label{figspecificheat}}
\end{figure}

While this certainly does not allow one to predict the
reliability of our HDL result for real, finite-$N_f$ QCD, it
should give an idea of the errors to expect at least.
Interestingly enough, in large-$N_f$ QCD
the higher-order corrections seem to increase
somewhat the range in $T/\mu$ where there is an excess of
the entropy over its ideal-gas value.

By the same token,
in Fig.~\ref{figSfullscd} we display the renormalization scale dependence
of the HDL result by varying the renormalization point by a factor
of 2 around a central value of $\bar\mu_{\rm MS}=2\mu$.

\subsection{Specific heat}

In Fig.~\ref{figspecificheat}, we finally evaluate our result
for the low-temperature specific heat at constant baryon density,
$\mathcal C_v$. As can be seen from Eq.~(\ref{CvLL}), this is a nonlinear
functional of the thermodynamic potential. However, for
$\mu\ll T$ and to leading order in $\g$ the anomalous contributions
provided by Eq.~(\ref{Pfull}) enter only through the logarithmic
derivative of the entropy, and the nonlinear terms in Eq.~(\ref{CvLL})
need only include the ideal-gas result. This defines the
results in Fig.~\ref{figspecificheat} labelled as ``strict''.

In order to have again an estimate of the uncertainties of
undetermined higher-order contributions, we also computed
$\mathcal C_v$ in a thermodynamically consistent manner directly
from the pressure, given by Eq.~(\ref{Pfull}) plus
the perturbative zero-temperature result to order $\g^2$. 
This has the slight deficiency of including
higher-order terms in the second term of Eq.~(\ref{CvLL})
beyond the accuracy of the first one.
The ``thermodynamically consistent''
result is displayed in Fig.~\ref{figspecificheat} by dashed
lines, and one can see that there is not much difference in the
region where $\mathcal C_v/\mathcal C_v^0$ is larger than one,

The results are given for three
different couplings. The lines marked ``QED'' correspond to
$\g=0.303$ or $\alpha_{QED}\approx 1/137$, and the results
for $\g=2,3$ correspond to $\alpha_s\approx 0.32,0.72$
in two-flavor QCD. (Recall that $\g^2\equiv g^2N_f/2$.)
While in QED the effect is tiny (the deviations
from the ideal-gas value have been enlarged by a factor of 20
in Fig.~\ref{figspecificheat} to make them more visible), in QCD we find that
there is an interesting range of $T/\mu$ where there is
a significant excess of the specific heat over its ideal-gas value.
whereas ordinary perturbation theory \cite{Kap:FTFT}
would have resulted in 
a low-temperature limit of
$\mathcal C_v/\mathcal C_v^0=1-2\alpha_s/\pi$.

According to Ref.~\cite{Rischke:2003mt},
the critical temperature for the color superconducting phase
transition may be anywhere between 6 and 60 MeV,
so with e.g.\ a quark chemical potential
of $\mu=500$ MeV the range $T/\mu \ge 0.012$ in Fig.~\ref{figspecificheat} 
might correspond to normal quark matter.

Thus, while the effect remains
small in QED, it seems conceivable that the anomalous terms in
the specific heat play a noticeable role in the thermodynamics of
proto-neutron stars,
in particular its cooling behavior in its earliest stages
before entering color superconductivity
\cite{Iwamoto:1980eb,Carter:2000xf,Wong:2004md}.

If color superconductivity leaves some quark matter components 
unpaired, the larger values of $\mathcal C_v/\mathcal C_v^0$
at smaller $T/\mu$ may also be relevant for %quark stars, or
neutron stars with a quark matter core.

\section{Summary} % Conclusions}

For temperatures much smaller than the chemical potential of
quarks (or electrons in the case of QED) we have computed
the leading contribution to the interaction
part of entropy and specific heat.
For temperatures smaller than the Debye mass $\propto \g\mu$,
the anomalous (non-Fermi-liquid) contributions become dominant.
As we have discussed at length, this effect can be viewed either
as a consequence of a logarithmic singularity of the fermion
self-energy at the Fermi surface caused by long-range quasi-static
magnetic interactions, or more directly as a contribution
of the (imaginary part of the) transverse gauge boson
propagator to the thermodynamic potential
when the hard fermion degrees of freedom are integrated out first.

This latter approach proved to be advantageous for a
systematic calculation beyond the well-known leading-log approximation.
We have obtained a hard-dense-loop resummed expression which
continuously interpolates between the more familiar plasmon
effect $\propto g^3$ coming from longitudinal Debye screened
gauge bosons and the non-Fermi-liquid effects coming from
only dynamically screened magnetic interactions. At temperatures
much smaller than the Debye mass, we have obtained a low-temperature
expansion starting with the well-known anomalous $T\ln T$ behaviour and
involving also fractional powers of $T$ in subleading terms.
The complete HDL-resummed result also contains
contributions which do not show up at any finite order of
the low-temperature expansion, being exponentially suppressed
by factors of $e^{-\omega_p/T}$ at small $T$, but which
become numerically important for intermediate temperatures.

Finally we have presented a numerical evaluation of our
HDL-resummed result together with its low-temperature expansion,
and we have compared with the exactly solvable large-$N_f$ limit
of QCD and QED. This comparison seems to indicate that our leading-order-in-$g$
result, which is equally applicable to finite-$N_f$ QCD,
is quite stable in the range of temperature where there is an
excess of entropy and specific heat over their respective
ideal-gas values. In QCD, where the coupling 
as well as the number of gauge bosons is much larger than
in QED, the deviation from
naive perturbation theory is appreciable for $T/\mu\lesssim 0.05$
and thus should be taken into account e.g.\ in studies of
thermodynamic properties of quark matter in (proto-)neutron stars.

\acknowledgments

We gratefully acknowledge extensive discussions with
Jean-Paul Blaizot and Urko Reinosa,
and a correspondence with Thomas Sch\"afer.
This work has
%A.~G.\ and A.~I.\ have 
been supported by the Austrian Science Foundation FWF,
project no. 16387-N08 and the Austrian-French exchange program
Amad\'ee of the \"OAD, project no. 16/2003.
%INT-04-01.

\appendix
\section{Calculation of $\tilde c$}

From the terms which are explicitly shown in Eq.~(\ref{cvh4}) 
we find the following contribution to the
coefficient of $b^{12}$ in the pressure, 
\begin{eqnarray}
  &&\!\!\!\!\!\!\!\!\!\!\!\!\tilde c_T^{(1)}={1\over810\pi^2}\Big[-36\pi^2+\nonumber\\
  &&\qquad(\pi^2-8)\Big(248-96\gamma_E-9\pi^2+48\ln(4\pi)
  +8640{\zeta^\prime(4)\over\pi^4}\Big)\Big]. \label{t42}
\end{eqnarray}
However, some of the terms in the integrand of Eq.~(\ref{cvh4})
that are formally of higher order than $b^{12}$ contribute
also at the order of $b^{12}$, because the $x$-integration
would be infrared divergent, were it not for the cutoff $x_{min}\propto b^6$.

Since $x_{min}$ depends on $\g$ only through $b$, 
and since we can drop terms in the integrand of Eq.~(\ref{cvh4})
involving $\g$ explicitly, 
it is sufficient to take the HDL self energy in the following.
Then the gluon self energy can be written as
\begin{equation}
  \Pi_T\simeq\Pi_T^{\rm HDL}\left({q_0\0q}\right)=
\g^2\mu^2H_T\left({b^2 y\over x^{1/3}}\right),
\end{equation}
with some function $H_T$. In the following we may
neglect the explicit $q^2$ and $q_0^2$ in Eq.~(\ref{cvh1}), because these two terms do not become singular for 
small $x$.\footnote{However,  $q^2$ and $q_0^2$ would have to be taken into account when summing
up the IR contributions to the coefficient of $T^6$, since for this coefficient also less IR singular terms are
important \cite{AGDiss}.} 
After expansion of the integrand with respect to $b$ we
then obtain integrals of the type
\begin{equation}
  b^6\int_{x_{min}}dx \left({b^2y\over x^{1/3}}\right)^n \sim {b^{12}y^3\over{n-3}}. \label{t44}
\end{equation}
Now we see clearly that from arbitrary powers of $b$ in the integrand we get contributions to the order
$b^{12}$ in $P_T$. The case $n=3$ corresponds to the term of order $b^{12}\ln b$, which we have 
evaluated already in Sect.~\ref{subsectT}.
For $n>3$ we can concentrate on the IR region and send to
upper integration limit in 
Eq.\ (\ref{t44}) to infinity.
%As we are interested only in contributions from the IR region, we may take $\infty$ as upper integration limit in 
%Eq.\ (\ref{t44}), since for $n>3$ we get then no 
%contribution from the upper integration limit. 
(The cases $n<3$ have been evaluated explicitly in Eq.\ (\ref{t42}).)
Furthermore we see from Eq.\ (\ref{t44}) that from the $y$ integration we always get a factor
\begin{equation}
  \int_0^\infty {dy\,y^3\over{e^y-1}}={\pi^4\over15}.
\end{equation}
The complete coefficient can thus be written as
\begin{equation}
  \tilde c_T=\tilde c_T^{(1)}
  -{\pi^4\over15}{1\over12\pi^4 b^6}\int_{4\pi b^6}^\infty dx \sum_{n=8}^\infty {b^n\over n!}\left(\left[
  {\partial^n\over\partial b^n}\arctan\left({\imag\Pi^{\rm HDL}_T\over\real\Pi^{\rm HDL}_T}\right)\right]\bigg|_{b=0,y=1}\right).
\end{equation}
This expression is in fact independent of $b$ (see Eq.\ (\ref{t44})). Therefore we may simply set $b=1$.
Summing up the (Taylor) series, we find after the substitution $x=4\pi z^3$
\begin{eqnarray}
  &&\tilde c_T=\tilde c_T^{(1)}-{\pi\over15}\int_1^\infty dz\Bigg({128+3\pi^4z^3-8\pi^2(2+3z^2)\over6\pi^3z}\nonumber\\
  &&\qquad+z^2
  \arctan\bigg[{\pi(1-z^2)\over 2z+(z^2-1)\ln\left({z+1\over z-1}\right)}\bigg]\Bigg).
\end{eqnarray}
From this expression we see that the complete HDL self energy is required for this coefficient (and not
only the expansion for small $q_0$, which is sufficient for the fractional powers and the logarithmic terms).
The remaining integral over the parameter $z$ can probably not be done analytically. Numerically
one readily finds
\begin{equation}
  \tilde c_T=-0.00178674305\ldots
\end{equation}

%%%%%%%%

The constant $\tilde c_L^{\rm II}$ in Eq.~(\ref{cvh10})
can be determined by summing up IR enhanced contributions in a completely analogous manner.
The result is 
\begin{eqnarray}\label{ctildeL}
  &&\tilde c_L^{\rm II}={1\over8640}\Big(3\pi^4-2\pi^2(12-\pi^2)(-17+6\gamma_E-6\ln(\pi)-540{\zeta^\prime(4)\over\pi^4}
  \Big)\nonumber\\
  &&\qquad-{\pi\over30}\int_1^\infty dz\Bigg({\pi(\pi^2+12(-1-z^2+z^3))\over24z}\nonumber\\
  &&\qquad+z^2\arctan\bigg[-{2z\over\pi}+{1\over\pi}\ln\left({z+1\over z-1}\right)\bigg]\Bigg) %\nonumber\\
  %&&\quad
\simeq 0.11902569216\ldots
\end{eqnarray}

%\bibliographystyle{prsty}
%\bibliography{tft,tftpr,qft,ar,books}

\end{document}